\documentclass[aps,pra,reprint,superscriptaddress,floatfix]{revtex4-2}
\usepackage{epsfig}
\usepackage{graphicx}
\usepackage{color}
\usepackage{bm}
\usepackage{amssymb,amsmath,euscript,esint,mathtools}
\usepackage{times}
\usepackage{amsthm}
\usepackage{amsfonts}
\usepackage[english]{babel}
\usepackage{epstopdf}
\usepackage{multirow,makecell,array}
\usepackage{cancel}
\usepackage{braket}
\usepackage[dvipsnames]{xcolor}

\definecolor{amendments}{rgb}{0.0, 0.0, 1.0}
\newcommand{\am}[1]{{\color{amendments}{#1}}}
\usepackage[colorlinks=true,linkcolor=blue,citecolor=blue,urlcolor=blue]{hyperref}
\renewcommand{\am}[1]{#1}

\begin{document}
\newcommand{\Br}[1]{(\ref{#1})}
\newcommand{\Eq}[1]{Eq.~(\ref{#1})}
\newcommand{\frc}[2]{\raisebox{1pt}{$#1$}/\raisebox{-1pt}{$#2$}}
\newcommand{\frcc}[2]{\raisebox{0.3pt}{$#1$}/\raisebox{-0.3pt}{$#2$}}
\newcommand{\frccc}[2]{\raisebox{1pt}{$#1$}\big/\raisebox{-1pt}{$#2$}}
\newcommand{\RNumb}[1]{\uppercase\expandafter{\romannumeral #1\relax}}

\title{Parity nonconservation in hyperfine transitions of hydrogen-like muonic atoms}
\author{D. V. Chubukov}
\email{dmitrybeat@gmail.com}
\affiliation{Department of Physics, St. Petersburg State University, 7/9 Universitetskaya Naberezhnaya, St. Petersburg 199034, Russia}
\author{M. I. Kuznetsov}
\affiliation{School of Physics and Engineering, ITMO University, Kronverkskiy 49, St. Petersburg 197101, Russia}
\author{I. A. Aleksandrov}
\affiliation{Department of Physics, St. Petersburg State University, 7/9 Universitetskaya Naberezhnaya, St. Petersburg 199034, Russia}
\affiliation{Ioffe Institute, Politekhnicheskaya Street 26, St. Petersburg 194021, Russia}
\author{A. V. Malyshev}
\affiliation{Department of Physics, St. Petersburg State University, 7/9 Universitetskaya Naberezhnaya, St. Petersburg 199034, Russia}
\affiliation{Petersburg Nuclear Physics Institute named by B.~P.~Konstantinov of National Research Center ``Kurchatov Institute'', Gatchina 188300, Russia}

\begin{abstract}
We investigate parity nonconservation (PNC) in the $2s_{1/2}(F=0)\leftrightarrow 2s_{1/2}(F=1)$ hyperfine transition of hydrogenlike muonic $^{13}\mathrm{C}^{5+}$ and $^{15}\mathrm{N}^{6+}$. Finite-nuclear-size effects and the Uehling potential are incorporated directly into the Dirac equation, whereas the leading kinematic-recoil and Fermi--Breit hyperfine corrections to the energy levels are treated to first order in perturbation theory. The spin-independent PNC amplitude is evaluated using hyperfine-resolved energy denominators and is proportional to the weak charge $Q_W$. The spin-dependent amplitude is proportional to $\kappa=\kappa_a+\kappa_Z$, where $\kappa_a$ and $\kappa_Z$ denote the nuclear-anapole and direct neutral-current couplings, respectively. For absorption we obtain the PNC degree $\mathcal P_{\rm abs}(^{13}\mathrm{C})=-5.9\times10^{-8}Q_W +1.3\times10^{-5}\kappa$ and $\mathcal P_{\rm abs}(^{15}\mathrm{N})=-8.8\times10^{-9}Q_W -8.1\times10^{-6}\kappa$. Nuclear-structure calculations indicate that $^{13}\mathrm{C}$ has enhanced relative sensitivity to direct $Z^0$-boson exchange, whereas the spin-dependent interaction in $^{15}\mathrm{N}$ is expected to be dominated by the nuclear-anapole contribution. In an idealized cavity-polarimetry model, a column density of order $10^{15}~\mathrm{cm^{-2}}$ is required; however, the production of metastable hydrogenlike muonic ions remains the principal experimental challenge.
\end{abstract}
 
\maketitle

\section{Introduction}\label{sec:intro}

Atomic parity nonconservation (PNC) provides a sensitive low-energy probe of the electroweak sector of the Standard Model (SM) and of possible interactions beyond it (``new physics'')~\cite{GingesFlambaum2004,Khriplovich1991,Safronova2018}. Since the discovery of weak neutral currents, parity-violating effects have been studied both theoretically and experimentally in a variety of atomic systems. Their interpretation requires accurate calculations because the weak lepton--nucleus interaction is short-ranged and probes the bound-particle wave function in the nuclear region. In heavy many-electron atoms this requires an accurate treatment of electron correlations, the Breit interaction, and quantum-electrodynamic (QED) radiative corrections. \am{Accurate calculations of the Cs PNC amplitude, including electron correlation, Breit, and QED effects, have reached quoted theoretical uncertainties at the sub-percent level~\cite{Shabaev2005,Porsev2009,Dzuba2012}.}

Atomic PNC has been investigated through several complementary observables, including Stark interference, circular dichroism, and optical rotation. The theoretical foundations and early experimental proposals were developed in Refs.~\cite{BouchiatBouchiat1974,Khriplovich1974}, while landmark observations were reported in Refs.~\cite{BarkovZolotorev1978,Wood1997}. Such measurements test the weak charge at low momentum transfer and constrain possible extensions of the electroweak sector, including additional neutral gauge bosons~\cite{MarcianoRosner1990,Langacker09}. The rapid increase of spin-independent PNC effects with nuclear charge has also motivated the study of simpler few-electron highly charged ions. Opposite-parity level crossings can provide an additional enhancement in these systems~\cite{Gor74,Bond18}, although no PNC effect in a highly charged ion has yet been observed.

\am{PNC in ordinary hydrogen and deuterium has been studied through weak
$2s$--$2p$ mixing and microwave transitions between metastable $2s$
components~\cite{HindsHughes1977,DunfordLewisWilliams1978,DunfordHolt2007}.
Experimental searches in atomic hydrogen have placed bounds on
parity-violating interactions~\cite{LevyWilliams1984}.
More recent proposals employ stimulated emission in an optical
cavity~\cite{RasorYost2020} and coherent control of the hydrogen
$2s$--$3s$ or $2s$--$4s$ transitions~\cite{LiDereviankoElliott2024}.
The earlier microwave E1--M1 interference schemes provide a close
methodological precedent for the hyperfine-transition approach
considered here.}

Muonic atoms provide another, qualitatively different class of simple systems with enhanced sensitivity to short-range interactions. The muon is about 207 times heavier than the electron, so that its Bohr radius is strongly reduced and its wave function substantially overlaps the nucleus. Muonic spectra are consequently highly sensitive
to nuclear charge distributions and polarizability, recoil, vacuum polarization, and other bound-state-QED effects~\cite{BorieRinker1982,GorringeHertzog2015,
AntogniniHagelsteinPascalutsa2022,Pachucki2024}. When the ordinary electronic shells are absent, a one-muon ion is at the same time an essentially one-particle atomic system and is free from the many-electron correlation uncertainties present in heavy neutral atoms.

The experimental capabilities for studying muonic atoms have advanced substantially in recent years. Laser spectroscopy of muonic hydrogen and helium by the CREMA collaboration at PSI has established precision spectroscopy of short-lived muonic systems and yielded benchmark determinations of nuclear-charge radii~\cite{Pohl2010,Antognini2013,Krauth2021,Antognini2021}. The program is being extended to the ground-state hyperfine splitting of muonic hydrogen \cite{Amaro2022}. The muX collaboration is developing precision x-ray spectroscopy of heavier muonic atoms
\cite{WautersKnecht2021}, while the MuSEUM collaboration is pursuing precision hyperfine spectroscopy of muonium and muonic helium at J-PARC~\cite{Strasser2025}. A dedicated feasibility study of atomic parity violation in muonic atoms has also been carried out at J-PARC~\cite{Kanda2022}. This experimental progress makes it
timely to reconsider PNC observables based on coherent light--matter interactions in metastable muonic states.

Parity violation in muonic atoms was investigated soon after the discovery of weak neutral currents. In Ref.~\cite{Bernabeu1974}, the strong sensitivity of the $2s\to1s$ transition in light muonic atoms was investigated. In Refs.~\cite{FeinbergChen1974,FeinbergChenHFS1974}, the parity-forbidden one-photon
$2s_{1/2}\to1s_{1/2}$ decay and the role of hyperfine interactions in this process were studied. The early proposals were reviewed in Ref.~\cite{MissimerSimons1985}. Subsequent studies considered the electroweak interpretation and possible nonstandard interactions~\cite{Langacker1991,Casalbuoni1991,Batell2011}, and direct radiative
population of the $2s$ state was proposed as a route to a parity-violating photon asymmetry~\cite{McKeenPospelov2012}. Nevertheless, no PNC effect in a muonic atom has yet been observed.

In the present work we consider the hyperfine transition $2s_{1/2}(F=0)\leftrightarrow2s_{1/2}(F=1)$ in hydrogen-like muonic atoms with nuclear spin $I=1/2$, for which the relatively small $2s_{1/2}$--$2p_{1/2}$ interval enhances the weak admixture. Since for $Z=6$ and $Z=7$ the $2s$ hyperfine interval lies in the optical or near-infrared region, we focus on $^{13}\mathrm C$ and $^{15}\mathrm N$. Throughout this work, a hydrogen-like muonic atom denotes a single negative muon bound to a bare nucleus, with no ordinary electrons. Therefore, the one-muon ions $^{13}\mathrm C$ and $^{15}\mathrm N$ carry net charges $+5$ and $+6$, respectively.

A negative muon captured by a nucleus initially occupies a highly excited state and then undergoes an Auger and radiative cascade~\cite{Ponomarev1973}. The cascade may remove most or all ordinary electrons, allowing the formation of one-muon ions with vacant electronic shells~\cite{Bacher1985}. A fraction of the captured muons reaches
the metastable $2s$ state. In muonic boron, an initial $2s$ population of a few percent was measured~\cite{Kirch1997}. Calculations show that the radiative decay of the $2s$ state of light one-muon ions is dominated by the two-photon $2s\to1s$ channel, with additional finite-nuclear-size-induced cascade contributions becoming important as $Z$ increases \cite{Knyazeva2022}. For the carbon and nitrogen systems, a simple mass- and charge-scaling estimate suggests the characteristic lifetime to be on the order of ten nanoseconds. Since complete isotope-specific decay calculations are not yet available, we use this lifetime in the feasibility estimates. 

Both spin-independent (SI) and spin-dependent (SD) PNC interactions contribute. For the hyperfine transition considered here, the contribution generated by the SI Hamiltonian is proportional to the nuclear weak charge $Q_W$ and vanishes if the hyperfine interaction is neglected. A nonzero contribution appears when the hyperfine-resolved
level energies are retained in the energy denominators. In the present work the hyperfine interaction is included in this diagonal form, while hyperfine-induced corrections to the wave functions are neglected. The SD contribution is described by the coupling constant $\kappa=\kappa_a+\kappa_Z$, where $\kappa_a$ and
$\kappa_Z$ denote the nuclear-anapole and direct $Z^0$-exchange contributions, respectively. The two nuclei considered provide complementary sensitivity within the SD sector. Nuclear calculations predict that direct $Z^0$-boson exchange is the largest contribution to $\kappa$ in $^{13}\mathrm C$ in the single-particle model and remains comparable to the anapole contribution in a no-core shell-model calculation. The SD interaction in $^{15}\mathrm N$ is instead predicted to be predominantly anapole driven~\cite{Hao2020}.

We calculate the contributions generated by the SI and SD PNC Hamiltonians using both a two-parameter Fermi and a uniformly-charged sphere models for the nuclear-charge distribution. The calculation includes the finite-nucleus Uehling potential, leading
kinematic recoil, and relativistic Fermi--Breit hyperfine shifts. Finally, we investigate a possible cavity-enhanced polarimetric measurement. Large effective optical paths and shot-noise-limited polarimetry have been demonstrated and analyzed in
Refs.~\cite{Baev1999,Bougas2012,Bougas2014,Durand2010}. The use of intracavity optical rotation with particle beams, including finite transit times, absorption, photon statistics, and systematic effects, has been developed in Refs.~\cite{Chubukov2017,Chubukov2019,Chubukov2021, Chubukov2023,Chekhovskoi2023}. These methods provide the basis for the feasibility analysis presented below. The purpose of this analysis is to identify the required column density and optical parameters and to separate fundamental sensitivity requirements from the present limitations of muonic-atom sources.

Throughout the paper we use relativistic units $\hbar=c=m_e=1$, where $m_e$ is the electron mass. The charge units are defined by $\alpha=e^2/(4\pi)$, where $e<0$ denotes the charge of negative leptons.

\section{Mechanism of parity nonconservation in hyperfine transitions} \label{sec:pnc}

\subsection{Estimate of the hyperfine splitting}

In the present work we focus on the hyperfine transition between the $F=0$ and $F=1$ components of the metastable $2s_{1/2}$ state in hydrogen-like muonic atoms with nuclear spin $I=1/2$. Important practical considerations are the stable nuclei and the transition frequency, which must lie in a range accessible to modern lasers.

For an order-of-magnitude estimate, we use the standard nonrelativistic Fermi expression for the hyperfine splitting of the $2s$ state in a hydrogen-like muonic atom with a pointlike nucleus with spin $I=1/2$~\cite{BetheSalpeter1957}:
\begin{equation}
\Delta E_{\mathrm{HFS}}(2s)
=
\frac{1}{3}\,
Z^3 \alpha^4\,
\frac{m_\mu^2}{m_p}\,
\frac{\mu_I}{\mu_N} ,
\label{eq:HFS_Fermi}
\end{equation}
where $\mu_I$ is the nuclear magnetic moment,  $\mu_N$ is the nuclear magneton; $m_\mu$ and $m_p$ are the muon and proton masses, respectively. Equation~(\ref{eq:HFS_Fermi}) is used here solely to estimate the typical frequency range of the $2s_{1/2}$ hyperfine transition. The corresponding estimated wavelengths for different stable isotopes are listed in Table~\ref{tab:HFS_lambda}.

\begin{table}[b]
\centering
\caption{Estimated wavelengths of the hyperfine transition $2s_{1/2}(F=0 \leftrightarrow F=1)$ in hydrogen-like muonic atoms with stable nuclei of spin $I=1/2$.
The estimates are based on the nonrelativistic Fermi formula for a point-like nucleus. The $\mu_I/\mu_N$ values are taken from Refs.~\cite{Radzig85,Stone05}.}
\label{tab:HFS_lambda}
\begin{ruledtabular}
\begin{tabular}{lccc}
Isotope & $Z$ & $\mu_I/\mu_N$ & $\lambda_{\mathrm{HFS}}$ (nm) \\
\hline
\noalign{\vskip 0.4ex}
$^{3}$He   & 2  & $-2.1275$ & $6512$ \\
$^{13}$C  & 6  & $+0.7024$ & $731$ \\
$^{15}$N  & 7  & $-0.2832$ & $1138$ \\
$^{19}$F  & 9  & $+2.6289$ & $58$ \\
$^{29}$Si & 14 & $-0.5553$ & $72$ \\
$^{31}$P  & 15 & $+1.1316$ & $29$ \\
$^{57}$Fe & 26 & $+0.0906$ & $70$ 
\end{tabular}
\end{ruledtabular}
\end{table}

Apart from isotope-to-isotope variations of the nuclear magnetic moment, the characteristic hyperfine wavelength decreases with the increase of $Z$. Table~\ref{tab:HFS_lambda} shows that the transitions in $^{13}\mathrm C$ and
$^{15}\mathrm N$ lie in the optical and near-infrared ranges, respectively.

\subsection{Parity-nonconserving interaction} \label{sec:pnc_hamiltonian}

The parity-nonconserving interaction between a lepton and the nucleus in an atom can be separated into nuclear-spin-independent (SI) and nuclear-spin-dependent (SD) contributions. The dominant part of the SI PNC interaction is described, in the one-boson-exchange approximation, by the effective Hamiltonian~\cite{GingesFlambaum2004,Khriplovich1991}
\begin{equation}
H_{\mathrm{PNC}}^{\mathrm{SI}}
=
-\frac{G_{\rm F}}{2\sqrt{2}}\, Q_W \gamma_5 \rho(\bf{r}),
\label{eq:H_SI}
\end{equation}
where  $G_{\rm F} \approx (1.02683 \times 10^{-5}) \hbar c (\hbar/m_p c)^2 \approx 3.04565\times10^{-12} \, \text{r.u.}$ is the Fermi coupling constant,
\begin{equation}
\gamma_5
=
-\begin{pmatrix}
0 & I_{2\times 2}\\
I_{2\times 2} & 0
\end{pmatrix}    
\end{equation}
is the Dirac matrix acting on the lepton four-component wave function, and $I_{2\times 2}$ is the two-by-two identity matrix. Here $\rho(\mathbf r)$ is the nuclear
density normalized to unity, $\int d\mathbf{r}\,\rho(\mathbf r)=1$. For the spherically
symmetric nuclear models used below, $\rho(\mathbf r)=\rho(r)$ with $r=|\mathbf r|$, so that
\begin{equation}
4\pi\int\limits_0^\infty r^2\rho(r)\,dr=1.
\label{eq:density_normalization}
\end{equation}
After averaging over the nuclear degrees of freedom, Hamiltonian~(\ref{eq:H_SI}) acts in the bound-muon space. The nuclear weak charge is defined by
\begin{equation}
 \label{Qw}
 Q_W = -N +Z (1-4\sin^2{\theta_W}),
\end{equation}
where $N$ and $Z$ are the numbers of neutrons and protons in a nucleus, respectively, and $\theta_W$ is the weak mixing angle. For the illustrative estimates below we use the tree-level value $\sin^2\theta_W\simeq0.23$~\cite{PDG2026}.

Nuclear-spin-dependent parity nonconservation is conventionally parameterized
by the effective Hamiltonian~\cite{GingesFlambaum2004,Khriplovich1991}
\begin{equation}
H^{\mathrm{SD}}_{\mathrm{PNC}}
=
\frac{G_{\rm F}}{\sqrt{2}}\,
\frac{\kappa}{I}\,
(\boldsymbol{\alpha}\cdot\mathbf{I})\,
\rho(\bf{r}).
\label{eq:H_SD}
\end{equation}
In the representation used here,
\begin{equation}
\boldsymbol{\alpha}
=
\begin{pmatrix}
0 & \boldsymbol{\sigma}\\
\boldsymbol{\sigma} & 0
\end{pmatrix},
\label{eq:alpha_matrices}
\end{equation}
where $\boldsymbol{\sigma}$ are the Pauli matrices. Here $\boldsymbol{\alpha}$ acts on the bound-muon spinor and $\mathbf I$ is the nuclear-spin operator. In the notation adopted throughout this work, $\kappa$ denotes the coupling constant of the SD interaction,
\begin{equation}
\kappa\equiv\kappa_a+\kappa_Z,
\label{eq:kappa_direct}
\end{equation}
where $\kappa_a$ is the nuclear-anapole contribution and $\kappa_Z$ is the direct neutral-current contribution.

The nuclear-structure-dependent contribution, $\kappa_a$, arises from the nuclear anapole moment generated by parity-violating nucleon--nucleon interactions. In the single-particle (SP) model, it is estimated as~\cite{Hao2020}
\begin{equation}
\kappa_a^{\rm SP}
\simeq
\frac{9}{10}
g_\nu
\frac{\alpha\mu_\nu}{m_p r_0}
A^{2/3}
\frac{K}{I+1}.
\label{eq:kappa_anapole}
\end{equation}
Here $A=N+Z$ is the nuclear mass number, and $\nu$ labels the unpaired valence nucleon; no summation over $\nu$ is implied. For $^{13}\mathrm C$ the valence nucleon is a neutron, $\nu = n$, whereas for $^{15}\mathrm N$ it is a proton, $\nu = p$. Also $\mu_p=2.793$ and $\mu_n=-1.913$ are their magnetic moments in nuclear magnetons, $r_0\simeq1.2~{\rm fm}$, and $K=(I+1/2)(-1)^{I-l_\nu+1/2}$.  The quantity $l_\nu$ is the orbital angular momentum. In the SP model, the nuclear spin is identified
with the total angular momentum of the unpaired valence nucleon, $I=j_\nu=l_\nu\pm1/2$, whose intrinsic spin is $1/2$. Thus, $I$ is half-integer within this approximation. The effective weak nucleon--nucleus couplings are poorly known; for the illustrative
single-particle estimates we use the central values $g_p=3.4$ and $g_n=0.9$ adopted in Ref.~\cite{Hao2020}. This contribution grows with nuclear size and dominates the SD PNC interaction in heavy nuclei.

The term $\kappa_Z$ represents the direct neutral-current interaction between the vector current of the bound lepton and the axial-vector current of the nucleus. In the SP model,
\begin{equation}
\kappa_Z^{\rm SP}
=
\frac{1/2-K}{I+1}C_2^{(\nu)},
\label{eq:kappa_Z}
\end{equation}
where $C_2^{(\nu)}$ is the coupling constant for the valence nucleon. At the tree level, for protons and neutrons it can be expressed as follows: 
\begin{equation}
C_2^{(n)} = -C_2^{(p)} = \frac{g_A}{2}
\left(1-4\sin^2\theta_W\right).
\label{eq:C2_atomic}
\end{equation}
Our convention is related to the scattering coefficients of Ref.~\cite{Hao2020} by $C_2^{(\nu)}=-C_{2\nu}^{\rm sc}$. For the two nuclei considered here, $I=1/2$, $g_A \approx 1.25$, and $K=1$, which gives
\begin{equation}
\kappa_Z(^{13}{\rm C})\simeq-0.017,
\qquad
\kappa_Z(^{15}{\rm N})\simeq+0.017.
\end{equation}
Unlike the anapole moment, $\kappa_Z$ does not exhibit any enhancement with $A$. Within the simple nuclear shell model, only the unpaired valence nucleon carrying the total nuclear spin $I$ contributes to $\kappa_Z$.

Representative values from Ref.~\cite{Hao2020} are summarized in Table~\ref{tab:kappa_models}. The anapole and direct $Z^0$-boson exchange contributions were evaluated in the SP model and the no-core shell (NCS) model.

\begin{table}[t]
\caption{
SD couplings from Ref.~\cite{Hao2020}. The values of $\kappa_a$ and $\kappa_Z$ are obtained in the single-particle (SP) and no-core shell (NCS) models. The last column gives the total SD coupling constant $\kappa=\kappa_a+\kappa_Z$.
}
\label{tab:kappa_models}
\begin{ruledtabular}
\begin{tabular}{c c r r r}
Nucleus & Model & $\kappa_a$ & $\kappa_Z$ & $\kappa$\\
\hline
\noalign{\vskip 0.4ex}
$^{13}\mathrm C$ & SP   & $-0.007$ & $-0.017$ & $-0.024$\\
                  & NCS & $-0.028$ & $-0.019$ & $-0.047$\\
$^{15}\mathrm N$ & SP   & $+0.044$ & $+0.017$ & $+0.061$\\
                  & NCS & $+0.088$ & $+0.015$ & $+0.103$
\end{tabular}
\end{ruledtabular}
\end{table}

Within the SD sector, direct $Z^0$-boson exchange is the largest individual contribution in $^{13}{\rm C}$ in the SP model and remains comparable to the
anapole term in the NCS model. In $^{15}{\rm N}$, the anapole contribution dominates in both models. Consequently, $^{13}{\rm C}$ retains enhanced relative sensitivity to the direct spin-dependent neutral current, although its interpretation remains nuclear-model dependent. A similar dominance of the direct neutral-current contribution within the SD sector was found for rovibrational transitions of ortho-H$_2$~\cite{Chubukov2020}.

It should be noted that the combined action of the SI weak interaction and the
hyperfine interaction may generate nuclear-spin-dependent corrections to the PNC amplitude as well. \am{A complete treatment of this contribution includes both diagonal hyperfine corrections to the energy denominators and hyperfine-induced corrections to the wave functions~\cite{Johnson03}. The combined weak--hyperfine amplitude is commonly parameterized by an effective spin-dependent coupling $\kappa_{\rm hf}$; this parametrization does not represent an additional fundamental PNC interaction. In the present calculation, the diagonal part is retained through the hyperfine-resolved level energies, while hyperfine-induced
corrections to the wave functions are neglected. The retained weak--hyperfine contribution is included in the term proportional to $Q_W$, while $\kappa=\kappa_a+\kappa_Z$ denotes the direct SD coupling.
In Sec.~\ref{sec:results}, we express this already included
weak--hyperfine contribution in terms of an equivalent coupling
$\kappa_{\rm hf}^{\rm diag}$ solely for comparison with the
direct SD contribution.}

Throughout this paper, the labels SI and SD specify which PNC Hamiltonian, Eq.~(\ref{eq:H_SI}) or Eq.~(\ref{eq:H_SD}), generates the corresponding contribution. For the hyperfine transition considered here, the contribution generated by $H_{\rm PNC}^{\rm SI}$ becomes nonzero only after the hyperfine interaction is included.

\subsection{Degree of parity violation}

We consider PNC effects in radiative transitions between hyperfine components of the same muonic state. The interference between the dominant magnetic-dipole (M1) amplitude and the weak-induced electric-dipole (E1) amplitude leads to a measurable PNC effect, characterized by the degree of parity nonconservation for this transition, defined as the relative difference in transition probabilities for opposite photon helicities. As a result, the absorption (or emission) rate can be written in the form
\begin{equation}
\label{eq:prob}
W = W_0
\left[
1 + \mathcal{P} \,
(\boldsymbol{s}_{\text{ph}} \cdot \boldsymbol{\nu})
\right],
\end{equation}
where $W_0$ is the total parity-conserving rate, and $\mathcal{P}$ is a dimensionless coefficient characterizing the degree of parity violation in the transition.  Here $\boldsymbol{s}_{\rm ph} =i\,\boldsymbol{\epsilon}\times\boldsymbol{\epsilon}^{*}$
is the photon spin, $\boldsymbol{\epsilon}$ is the photon polarization,  $\boldsymbol{\nu}=\mathbf k/|\mathbf k|$ is the photon propagation direction, and $\boldsymbol{s}_{\rm ph}\cdot\boldsymbol{\nu}=\pm1$ is the photon helicity that changes sign under spatial inversion and therefore represents a direct manifestation of parity nonconservation.

To obtain the expression for the PNC degree, we consider the interaction of an atom with a monochromatic electromagnetic field. We distinguish the hyperfine-coupled states of the muon--nucleus system from the corresponding bound-muon states. The former are denoted by
\begin{equation}
|\gamma J;FM\rangle
\equiv
|(\gamma J\,I)FM\rangle,
\end{equation}
whereas the bound-muon states before coupling to the nuclear spin are written as $|\gamma Jm_J\rangle$. Here and in what follows $\gamma$ denotes all remaining bound-muon quantum numbers; $F$ and $M$ are the total angular momentum and its projection, respectively; $J$ and $m_J$ are the total muonic angular momentum and its projection, respectively. Accordingly, a reduced matrix element carrying an $F$ label refers to the
hyperfine-coupled space, whereas a reduced matrix element carrying only $J$ refers to the bound-muon space.  In particular, the initial and final states are
\begin{equation}
|i\rangle=|\gamma_iJ_i;F_iM_i\rangle,
\qquad
|f\rangle=|\gamma_fJ_f;F_fM_f\rangle.
\end{equation}
For the transition considered here, $\gamma_iJ_i=\gamma_fJ_f=2s_{1/2}$ and $J_i=J_f=1/2$. The transition probability between initial and final
states is determined by the transition amplitude
\begin{equation}
\mathcal{M}_{fi} = \langle f | \hat{H}_{\mathrm{int}} | i \rangle ,
\end{equation}
where $\hat{H}_{\mathrm{int}}$ is the interaction Hamiltonian with the radiation
field. As noted above, for hyperfine transitions within the same relativistic state,
the leading parity-allowed contribution is of magnetic-dipole type. In the presence of weak interactions, opposite-parity states are admixed to the initial and final states.
As a result, an additional parity-violating electric-dipole amplitude appears. The total transition amplitude can therefore be written as
\begin{equation}
\mathcal{M}_{fi}
=
\mathcal{M}_{fi}^{(\mathrm{M1})}
+
\mathcal{M}_{fi, \mathrm{PNC}}^{(\mathrm{E1})}+\dots.
\end{equation}

One can expand the interaction Hamiltonian in electromagnetic multipoles and keep the leading contributions relevant for our transition. Note that the electric-quadrupole (E2) amplitude does not contribute to the transition considered here since it is forbidden by angular-momentum conservation. For the hyperfine transitions considered here, $\omega_0r\ll1$, so retardation corrections of relative order $O[(\omega_0r)^2]$ can be neglected. The relevant operators for a photon with wave vector $\mathbf k$ and polarization vector $\boldsymbol{\epsilon}$ are
\begin{align}
\hat{H}_{\mathrm{E1}} &= -\omega_0\left(\boldsymbol{\epsilon} \cdot \mathbf{d}\right), \label{HE1} \\
\hat{H}_{\mathrm{M1}} &= -[\mathbf k \times \boldsymbol{\epsilon}] \cdot \boldsymbol{\mu}.
\label{HM1}
\end{align}
Here $\mathbf d$ and $\boldsymbol{\mu}$ are the internal electric- and magnetic-dipole operators of the muon--nucleus system, respectively. Their explicit forms are given below. The multipole operators must be constructed from the total electromagnetic current of the muon--nucleus system. The usual expansion of $\boldsymbol{\alpha}\cdot\boldsymbol{\epsilon}\exp(i\mathbf k\cdot\mathbf r)$ gives the muonic part of the interaction in the fixed-nucleus approximation. For a nucleus of finite mass, the internal E1 operator acquires the center-of-mass correction displayed below, while the nuclear magnetization current provides the nuclear M1 contribution.

The electric dipole moment is $e\mathbf r_\mu-Ze\mathbf r_{\rm nuc}$. After separating the center-of-mass coordinate, its internal part is $e\eta_E\mathbf r=-|e|\eta_E\mathbf r$. We therefore define the internal electric-dipole operator as
\begin{equation}
\mathbf d
=
e\eta_E\mathbf r,
\qquad
\eta_E=
\frac{M+Zm_\mu}{M+m_\mu},
\label{eq:E1_internal}
\end{equation}
where $M$ is the nuclear mass. The factor $\eta_E$ approaches unity in the infinite-nuclear-mass limit.

The magnetic-dipole operator relevant to the hyperfine transition is the sum of the muonic and nuclear contributions:
\begin{equation}
\boldsymbol{\mu}=\boldsymbol{\mu}_\mu+\boldsymbol{\mu}_{\rm nuc},
\label{eq:M1_total_operator}
\end{equation}
where
\begin{equation}
\boldsymbol{\mu}_\mu=\frac{e}{2}[\mathbf r\times\boldsymbol{\alpha}],
\qquad
\boldsymbol{\mu}_{\rm nuc}=\frac{\mu_I}{I}\mathbf I.
\label{eq:M1_components}
\end{equation}

The atomic states in the presence of parity nonconservation can be written as
\begin{equation}
| \widetilde{a} \rangle=| a \rangle+\sum_n\frac{\langle n | \hat{H}_{\mathrm{PNC}} | a \rangle}{E_a - E_n}| n \rangle ,
\end{equation}
where the sum runs over states $|n\rangle$ of opposite parity. As a result, the parity-violating E1 amplitude arises in second-order perturbation theory and can be written as
\begin{align}
\mathcal M_{fi,\mathrm{PNC}}^{(\mathrm{E1})}
&=
\sum_n
\frac{
\langle f|\hat H_{\mathrm{E1}}|n\rangle
\langle n|\hat H_{\mathrm{PNC}}|i\rangle
}{
E_i-E_n
}
\nonumber \\
{}&+
\sum_n
\frac{
\langle f|\hat H_{\mathrm{PNC}}|n\rangle
\langle n|\hat H_{\mathrm{E1}}|i\rangle
}{
E_f-E_n
}.
\end{align}

According to the time-reversal invariance of the theory, the PNC matrix elements are purely imaginary~\cite{Khriplovich1991}. After angular reduction, we therefore write
\begin{equation}
 \langle a|\hat H_{\rm PNC}|b\rangle
 =
 iS_{ab}\delta_{F_aF_b}\delta_{M_aM_b},
\end{equation}
where the additional angular-momentum selection rules depend on whether the SI or SD interaction is considered. Then, we follow the general approach developed in Sec.~5.2 of Ref.~\cite{Bond11}, adapting it to transitions between hyperfine components. The transition probability is proportional to the squared modulus of the total amplitude,
\begin{align}
W &\propto |\mathcal{M}_{fi}|^2
\approx
|\mathcal{M}^{(\mathrm{M1})}_{fi}|^2 \nonumber \\
{}&+ 2 \, \mathrm{Re}
\left(
\mathcal{M}^{(\mathrm{M1})\,*}_{fi}
\mathcal{M}^{(\mathrm{E1})}_{fi,\mathrm{PNC}}
\right).
\end{align}
Terms quadratic in the PNC-induced E1 amplitude, $|\mathcal M_{fi,\rm PNC}^{(\rm E1)}|^2$, have been neglected. The second term represents the interference of the parity-conserving M1 amplitude with the parity-violating E1 amplitude and gives rise to PNC effects. After applying the Wigner--Eckart theorem and summation over projections $M_i$ and $M_f$, the interference term acquires a characteristic angular dependence and is proportional to  the pseudoscalar $(\boldsymbol{s}_{\text{ph}} \cdot \boldsymbol{\nu}) = \pm 1$. Thus, \Eq{eq:prob} is justified. The degree of parity nonconservation $\mathcal{P}$ is determined by the ratio of the PNC-induced E1 amplitude to the dominant M1 amplitude:
\begin{widetext}
\begin{equation}
\label{eq:PNC_degree}
\mathcal{P} =  \frac{2\sum_n \bigg( \frac{\langle \gamma_fJ_f ; F_f \|  \boldsymbol{d} \| \gamma_n J_n ; F_i \rangle S_{ni}}{E_i-E_{nJ_nF_i}}  +  \frac{S_{fn} \langle \gamma_nJ_n; F_f \|  \boldsymbol{d} \| \gamma_i J_i;F_i \rangle }{E_f-E_{nJ_nF_f}}
  \bigg)}{\langle \gamma_fJ_f; F_f \|  \boldsymbol{\mu}\| \gamma_i J_i ;F_i \rangle} \equiv  \frac{2\langle \gamma_f J_f; F_f \| \boldsymbol{d}^{\text{PNC}} \| \gamma_i J_i;F_i \rangle}{\langle \gamma_f J_f; F_f \| \boldsymbol{\mu}\| \gamma_i J_i; F_i \rangle},
\end{equation}
where $\langle\cdots\|\boldsymbol{d}\|\cdots\rangle$ and $\langle\cdots\|\boldsymbol{\mu}\|\cdots\rangle$ are the reduced electric- and magnetic-dipole matrix elements, respectively. 

\am{The general angular reduction is given in
Appendix~\ref{app:angular}. For $I=J=1/2$ and the ordered
transition $F_i=0\to F_f=1$, the SI amplitude is}
\begin{equation}
\langle \gamma J; F_f || \boldsymbol{d}^{\rm PNC}_{\rm SI} || \gamma J; F_i \rangle = -\frac{G_{\rm F}Q_W}{4} \sum_n
\langle \gamma J ||  \boldsymbol{d} || \gamma_nJ_n\rangle
\langle \gamma_nJ_n | (\gamma_5/i)\, \rho(r) | \gamma J\rangle
  \left(\frac{1}{E_i-E_{nJ_nF_i}}  - \frac{1}{E_f-E_{nJ_nF_f}}\right).
\label{eq:SI_PNC3}
\end{equation}
If the hyperfine interaction is neglected, the two denominators coincide and the contribution generated by $H_{\rm PNC}^{\rm SI}$
vanishes for the present transition. The leading nonzero contribution
is linear in the hyperfine interaction. In the diagonal-hyperfine
approximation adopted here, this effect is retained through the level
energies, whereas hyperfine-induced corrections to the wave functions
are omitted. Expanding the denominator difference to first order in
the hyperfine interaction gives the additional suppression factor for
the first admixed state ($2p$) of the order
$\left[\Delta E_{\rm HFS} (2s)
-\Delta E_{\rm HFS}(2p_{1/2})\right]/\Delta E (2s-2p)
\sim 10^{-2}$.

\am{The corresponding SD amplitude is}
\begin{align}
\langle \gamma J; F_f || \boldsymbol{d}_{\rm PNC}^{\rm SD} || \gamma J; F_i \rangle &=
-\frac{G_{\rm F}\kappa}{2\sqrt{6}} \sum_n \langle \gamma J ||  \boldsymbol{d} || \gamma_n J_n\rangle
\langle \gamma_n J_n || (\boldsymbol{\alpha}/i) \rho (r)|| \gamma J\rangle \nonumber \\
{}&\times \left[ s(J_n) \frac{1}{E_i-E_{nJ_nF_i}} + k(J_n) \frac{1}{E_f-E_{nJ_nF_f}} \right],
\label{eq:SD_PNC2}
\end{align}
\end{widetext}
where $s(1/2)=3$, $s(3/2)=0$, $k(1/2)=1$, and $k(3/2)=2$. \am{The sum runs over odd-parity states with $J_n=1/2,3/2$.} For consistency with the SI calculation, the same hyperfine-resolved level energies are used in the SD denominators. Unlike in the SI contribution, however, the hyperfine splitting is not required for the SD amplitude to be nonzero.

For $I=J=1/2$ and for the ordered transition $F_i=0\to F_f=1$, the total reduced magnetic-dipole matrix element is
\begin{equation}
\begin{split}
\langle \gamma J; F_f
\Vert\boldsymbol{\mu}\Vert
\gamma J; F_i\rangle
={}&
\frac{1}{\sqrt{2}}
\langle \gamma J
\Vert\boldsymbol{\mu}_{\mu}\Vert
\gamma J\rangle
-\sqrt{3}\,\mu_I .
\end{split}
\label{eq:M1_total_reduced}
\end{equation}
Defining the tabulated dimensionless magnetic moment as
\begin{equation}
\widetilde{\mu}_I
\equiv
\frac{\mu_I}{\mu_N},
\end{equation}
the dimensional quantity entering
Eq.~(\ref{eq:M1_total_reduced}) is
\begin{equation}
\mu_I
=
\widetilde{\mu}_I
\frac{|e|}{2m_p}.
\end{equation}

The amplitudes below are defined for the ordered transition $F_i=0\to F_f=1$. Under reversal of the transition, the reduced magnetic-dipole matrix element changes sign, whereas the real coefficient conventionally associated with the purely imaginary PNC electric-dipole amplitude remains unchanged. Consequently, $\mathcal P_{1\to0}=-\mathcal P_{0\to1}$. Here and below, $F_g$ and $F_e$ denote the lower and upper
hyperfine components, respectively, participating in the physical absorption transition. For $^{13}\mathrm C$, the absorption transition is $F_g=0\to F_e=1$, whereas for $^{15}\mathrm N$ it is $F_g=1\to F_e=0$.

\section{Calculations} \label{sec:calculations}

The numerical calculation was performed by solving the radial Dirac equation for a muon in the central potential
\begin{equation}
V(r)=V_C(r)+V_U(r),
\end{equation}
where $V_C$ is the finite-nucleus electrostatic potential and $V_U$ is the Uehling potential described in detail below. The four-component wave function is represented as
\begin{equation}
\psi_{n\varkappa m}(\mathbf r)
=
\frac{1}{r}
\begin{pmatrix}
G_{n\varkappa}(r)\,
\Omega_{\varkappa m}(\hat{\mathbf r})
\\[6pt]
i\,F_{n\varkappa}(r)\,
\Omega_{-\varkappa m}(\hat{\mathbf r})
\end{pmatrix},
\label{eq:Dirac_wavefunction}
\end{equation}
where $\Omega_{\pm\varkappa m}(\hat{\mathbf r})$ are the spinor spherical harmonics, $G_{n\varkappa}(r)$ and $F_{n\varkappa}(r)$ are the large and small radial components of the Dirac wave function, respectively, with the standard normalization condition
\begin{equation}
\int\limits_0^\infty
\left[
G_{n\varkappa}^2(r)+F_{n\varkappa}^2(r)
\right]dr=1 .
\end{equation}
Here $\varkappa=(-1)^{J+l+1/2}(J+1/2)$ is the relativistic angular quantum number. With this convention the radial equations have the form
\begin{align}
\frac{dG}{dr}
&=
-\frac{\varkappa}{r}G+
\left[E+m_\mu-V(r)\right]F,
\\
\frac{dF}{dr}
&=
\frac{\varkappa}{r}F-
\left[E-m_\mu-V(r)\right]G .
\end{align}

In the present calculation, the charge, weak-charge, and spin-dependent nuclear densities are represented by the same normalized radial profiles. We use two profiles: a uniformly charged sphere and a two-parameter Fermi distribution. This common-density approximation allows us to isolate the sensitivity to the radial shape; it does not include possible differences among the proton, neutron, and spin densities. The common-density approximation is sufficient for the exploratory estimate presented here but is not intended as a precision nuclear-structure treatment.

The electrostatic Coulomb potential is written as
\begin{equation}
V_C(r)=-\alpha Z\,v_r(r).
\end{equation}
For the uniformly charged sphere,
\begin{equation}
v_r(r)=
\begin{cases}
\displaystyle
\frac{3/2-(r/R)^2/2}{R}, & r\leq R,\\[2mm]
\displaystyle
\frac{1}{r}, & r>R,
\end{cases}
\end{equation}
where
\begin{equation}
R=\sqrt{\frac{5}{3}}\,r_{\rm rms}.
\end{equation}
Here $r_{\rm rms}=\sqrt{\langle r^2\rangle}$ is the root-mean-square (RMS) nuclear charge radius. The corresponding normalized density is
\begin{equation}
\rho_{\rm sph}(r)=
\begin{cases}
\displaystyle \frac{3}{4\pi R^3}, & r\leq R,\\[2mm]
0, & r>R.
\end{cases}
\end{equation}

For the two-parameter Fermi model,
\begin{align}
\rho_{\rm F}(r)
&=
\frac{\rho_0}{1+\exp[(r-c)/a]},
\\
a&=\frac{t}{4\ln 3},
\qquad
t=2.3~{\rm fm}.
\label{eq:Fermi_density}
\end{align}
For a general spherically symmetric normalized density, the radial factor entering the electrostatic potential is
\begin{equation}
v_r(r)=4\pi\left[
\frac{1}{r}\int\limits_0^r r'^2\rho(r')\,dr'
+
\int\limits_r^\infty r'\rho(r')\,dr'
\right].
\label{eq:finite_nucleus_potential}
\end{equation}
Equation~\eqref{eq:finite_nucleus_potential} is used with $\rho=\rho_{\rm F}$ for the Fermi model and reproduces the piecewise expression above for $\rho=\rho_{\rm sph}$. The half-density radius $c$ and normalization constant $\rho_0$ are determined from
\begin{equation}
4\pi\int\limits_0^\infty r^2\rho_{\rm F}(r)\,dr=1,
\quad
4\pi\int\limits_0^\infty r^4\rho_{\rm F}(r)\,dr
=
r_{\rm rms}^2.
\end{equation}
For the two systems considered here, the input RMS charge radii are
\begin{align}
r_{\rm rms}(^{13}{\rm C}) &=2.4464(45)~{\rm fm}, \label{eq:rms_radii_C} \\
r_{\rm rms}(^{15}{\rm N})&=2.6058(80)~{\rm fm}.
\label{eq:rms_radii_N}
\end{align}
The $^{13}{\rm C}$ value is taken from the recent isotope-shift determination of Ref.~\cite{Muller2025}, while the $^{15}{\rm N}$ value is the adopted value of Ref.~\cite{AngeliMarinova2013}.  The Fermi distribution is used for the reference numerical results below. The sphere calculation is retained as an alternative profile and is used to quantify the density-shape sensitivity.

The finite-nucleus Uehling potential was incorporated directly into the Dirac Hamiltonian. For a spherically symmetric finite charge distribution it is evaluated as~\cite{Serber1935,Uehling1935}
\begin{align}
V_U(r)
&=
-\frac{2Z\alpha^2}{3r}
\int\limits_0^\infty dr'\,r'\rho(r') \nonumber
\\
{}&\times
\left[
K_0\!\left(2|r-r'|\right)
-
K_0\!\left(2(r+r')\right)
\right],
\end{align}
where
\begin{equation}
K_0(x)=
\int\limits_1^\infty dt\,
\mathrm{e}^{-xt}
\left(
\frac{1}{t^3}
+
\frac{1}{2t^5}
\right)
\sqrt{t^2-1}.
\end{equation}
The value of $V_U(r)$ at the origin was evaluated by taking the finite-nucleus
limit separately in order to avoid numerical cancellation.

The Dirac equation was solved with the physical muon mass $m_\mu=206.7682843\,m_e$~\cite{Mohr2012}. The leading kinematic recoil contribution was evaluated after solving the Dirac equation as
\begin{equation}
\Delta E_{\rm rec}^{\rm kin}
=
\frac{1}{2M}
\langle \mathbf p^2\rangle ,
\label{eq:recoil_kin}
\end{equation}
where $M$ is the nuclear mass. The term in Eq.~(\ref{eq:recoil_kin}) coincides with the
leading contribution in the rigorous QED treatment of nuclear recoil in muonic atoms~\cite{YerokhinOreshkina2023}. It was shown in Ref.~\cite{YerokhinOreshkina2023} that this leading term already provides a very reasonable approximation to the
complete recoil correction, with the difference decreasing toward lower $Z$ and more weakly bound states. In the numerical evaluation we use $M=Am_p$, where
$m_p=1836.15267245\,m_e$~\cite{Mohr2012}, both in the leading recoil correction and in the finite-mass factor $\eta_E$ of Eq.~(\ref{eq:E1_internal}). Replacing this approximation by the experimental nuclear masses would produce a change well below the
accuracy considered here.

A recent self-consistent analysis of atomic effects in bound-muon decay showed that, for muonic carbon, the finite-nuclear-size, Uehling, and leading recoil corrections to
the bound-muon energy are comparable in scale, whereas the Wichmann--Kroll contribution is much smaller~\cite{Kaygorodov2026}. Although the observable and the reference state considered in that work differ from the present $2s$ hyperfine transition, these results support the inclusion of finite nuclear size, Uehling vacuum polarization,
and recoil already at the baseline level.

A recent systematic framework for muonic atoms in the range $3\le Z\lesssim30$ includes higher-order QED, recoil, and nuclear-polarization effects~\cite{Rathi2026}. In this region vacuum polarization remains the leading QED contribution, whereas the
self-energy contribution is comparatively smaller. These developments provide a natural basis for a future precision refinement of the present calculation.

The radial Dirac equations were solved to high numerical precision using a
fourth-order Runge--Kutta shooting method. Convergence tests showed that the numerical uncertainties of the energies and radial matrix elements are negligible compared with the physical uncertainties discussed below. 

The reference state in the matrix-element calculations was $2s_{1/2}$, and the positive-energy bound intermediate states were $np_{1/2}$ and $np_{3/2}$. In the reference
calculation, the discrete bound-state sum was evaluated explicitly up to $n=10$. The lowest $p_{1/2}$ state dominates because of its small energy denominator and its large
density-weighted overlap with the nuclear region, while the higher bound-state contributions decrease rapidly. The numerical quadratures in the reference calculation were performed using a six-point Gauss--Legendre rule on each grid interval.

The convergence of the intermediate-state sum was tested independently using finite B-spline representations of the Dirac spectrum, covering both the positive- and
negative-energy continua. Two different realizations were employed for this aim, see Refs.~\cite{Johnson1988, Shabaev2004}. Calculations with different basis sizes gave consistent results. The full-spectrum results obtained with the different
basis constructions and sizes agree to better than $10^{-4}\%$. The resulting contributions beyond the explicitly summed $n\leq10$ bound states were found
to be negligible at the present level of accuracy.

\am{The hyperfine-resolved energy denominators in Eqs.~(\ref{eq:SI_PNC3}) and (\ref{eq:SD_PNC2}) are evaluated using the relativistic Fermi--Breit expression~\cite{BetheSalpeter1957} for a point nuclear magnetic dipole}. The finite charge distribution of the nucleus is included in the Dirac wave functions, whereas the finite magnetization distribution, i.e. the Bohr--Weisskopf correction, is neglected. In fact, the dominant contribution to the $2s_{1/2}$--$2p_{1/2}$ interval in muonic atoms comes from the finite nuclear-charge distribution and vacuum polarization. The Bohr--Weisskopf correction is expected to have a much smaller relative effect on the dominant denominator, and hence on the leading SD amplitude, than on the hyperfine intervals themselves. The contribution generated by the SI PNC Hamiltonian depends on the difference between the hyperfine corrections to the $2s_{1/2}$ and intermediate $np_{1/2}$ states and is therefore particularly sensitive to the finite magnetization distribution. Its possible contribution is included phenomenologically in the working allowance assigned to $C_{Q_W}$ below. Hyperfine effects are included through diagonal matrix elements in the level energies entering the denominators. Hyperfine-induced corrections to the wave functions are not included. This treatment defines the diagonal-hyperfine approximation used throughout the present work. \am{The radial expressions for the M1, E1, and weak matrix elements are collected in Appendix~\ref{app:radial}.}

\section{Results} \label{sec:results}

\subsection{Energy levels and hyperfine structure}

The energies include the rest energy of the muon unless the binding energy
\begin{equation}
\varepsilon_{n\varkappa F}=E_{n\varkappa F}-m_\mu
\end{equation}
is explicitly used. The hyperfine-resolved energies entering the denominators in Eqs.~(\ref{eq:SI_PNC3}) and (\ref{eq:SD_PNC2}) were constructed as
\begin{equation}
\varepsilon_{n\varkappa F}
=
\varepsilon^{\rm D}_{n\varkappa}
+
\Delta E^{\rm rec}_{n\varkappa}
+
\Delta E^{\rm FB}_{n\varkappa F},
\label{eq:energy_decomposition}
\end{equation}
where $\varepsilon^{\rm D}_{n\varkappa}=E^{\rm D}_{n\varkappa}-m_\mu$
is the Dirac binding energy obtained for the finite nuclear charge
distribution with the finite-nucleus Uehling potential incorporated into the Dirac Hamiltonian, $\Delta E^{\rm rec}_{n\varkappa}$ is the leading kinematic recoil correction, and $\Delta E^{\rm FB}_{n\varkappa F}$ is the relativistic
Fermi--Breit hyperfine shift. The resulting binding energies for the levels that enter the dominant $n=2$ and $n=3$ terms of the PNC amplitudes are listed in Table~\ref{tab:energies}.  The last column of Table~\ref{tab:energies} gives the total binding energy obtained by summing these three contributions.

\begin{table*}[t]
\caption{
Contributions to the binding energies $\varepsilon_{n\varkappa F}=E_{n\varkappa F}-m_\mu$ in relativistic units for the reference Fermi nuclear-density distribution. The last column gives the sum of the Dirac, recoil, and Fermi--Breit contributions. Only the hyperfine components relevant to the present calculation are listed. Additional digits are retained to reproduce the energy denominators and should not be interpreted as the physical accuracy of the levels.
}
\label{tab:energies}
\begin{ruledtabular}
\begin{tabular}{c c c r r r r}
Isotope &
State &
$F$ &
$\varepsilon^{\rm D}_{n\varkappa}$ &
$\Delta E^{\rm rec}_{n\varkappa}$ &
$\Delta E^{\rm FB}_{n\varkappa F}$ &
$\varepsilon_{n\varkappa F}$
\\
\hline
\noalign{\vskip 0.4ex}
$^{13}\mathrm C$
& $2s_{1/2}$ & 0
& $-4.95729773\times10^{-2}$
& $ 4.29506\times10^{-4}$
& $-2.25532\times10^{-6}$
& $-4.91457263\times10^{-2}$
\\
& $2s_{1/2}$ & 1
& $-4.95729773\times10^{-2}$
& $ 4.29506\times10^{-4}$
& $ 7.51775\times10^{-7}$
& $-4.91427193\times10^{-2}$
\\
& $2p_{1/2}$ & 0
& $-4.96269916\times10^{-2}$
& $ 4.31062\times10^{-4}$
& $-8.41137\times10^{-7}$
& $-4.91967708\times10^{-2}$
\\
& $2p_{1/2}$ & 1
& $-4.96269916\times10^{-2}$
& $ 4.31062\times10^{-4}$
& $ 2.80379\times10^{-7}$
& $-4.91956493\times10^{-2}$
\\
& $2p_{3/2}$ & 1
& $-4.96031075\times10^{-2}$
& $ 4.30440\times10^{-4}$
& $-2.79510\times10^{-7}$
& $-4.91729475\times10^{-2}$
\\
& $3p_{1/2}$ & 0
& $-2.20447260\times10^{-2}$
& $ 1.91329\times10^{-4}$
& $-2.49057\times10^{-7}$
& $-2.18536464\times10^{-2}$
\\
& $3p_{1/2}$ & 1
& $-2.20447260\times10^{-2}$
& $ 1.91329\times10^{-4}$
& $ 8.30190\times10^{-8}$
& $-2.18533143\times10^{-2}$
\\
& $3p_{3/2}$ & 1
& $-2.20376560\times10^{-2}$
& $ 1.91145\times10^{-4}$
& $-8.27700\times10^{-8}$
& $-2.18465942\times10^{-2}$
\\[1mm]
\hline
\noalign{\vskip 0.4ex}
$^{15}\mathrm N$
& $2s_{1/2}$ & 0
& $-6.74352750\times10^{-2}$
& $ 5.05927\times10^{-4}$
& $ 1.40257\times10^{-6}$
& $-6.69279458\times10^{-2}$
\\
& $2s_{1/2}$ & 1
& $-6.74352750\times10^{-2}$
& $ 5.05927\times10^{-4}$
& $-4.67522\times10^{-7}$
& $-6.69298159\times10^{-2}$
\\
& $2p_{1/2}$ & 0
& $-6.75763950\times10^{-2}$
& $ 5.09058\times10^{-4}$
& $ 5.39515\times10^{-7}$
& $-6.70667971\times10^{-2}$
\\
& $2p_{1/2}$ & 1
& $-6.75763950\times10^{-2}$
& $ 5.09058\times10^{-4}$
& $-1.79838\times10^{-7}$
& $-6.70675165\times10^{-2}$
\\
& $2p_{3/2}$ & 1
& $-6.75321315\times10^{-2}$
& $ 5.08059\times10^{-4}$
& $ 1.79088\times10^{-7}$
& $-6.70238936\times10^{-2}$
\\
& $3p_{1/2}$ & 0
& $-3.00140632\times10^{-2}$
& $ 2.25887\times10^{-4}$
& $ 1.59733\times10^{-7}$
& $-2.97880169\times10^{-2}$
\\
& $3p_{1/2}$ & 1
& $-3.00140632\times10^{-2}$
& $ 2.25887\times10^{-4}$
& $-5.32443\times10^{-8}$
& $-2.97882299\times10^{-2}$
\\
& $3p_{3/2}$ & 1
& $-3.00009643\times10^{-2}$
& $ 2.25591\times10^{-4}$
& $ 5.30296\times10^{-8}$
& $-2.97753204\times10^{-2}$
\end{tabular}
\end{ruledtabular}
\end{table*}

General calculations and tabulations of energy levels and higher-order corrections in muonic atoms can be found in Refs.~\cite{BorieRinker1982,RinkerSteffen1977}. Muonic x-ray transitions in carbon isotopes, including precise measurements of the $2p$--$1s$ transition in $^{13}\mathrm C$, were investigated in Refs.~\cite{Schaller1982,deBoer1985}. However, to the best of our knowledge,
hyperfine-resolved $2s$, $2p$, and $3p$ energies for the specific hydrogen-like muonic systems considered here have not previously been tabulated with the same set of finite-size, Uehling, recoil, and Fermi--Breit contributions.

Defining the hyperfine interval as
\begin{equation}
\Delta E_{\rm HFS}
=
E(2s_{1/2},F=1)-E(2s_{1/2},F=0),
\end{equation}
the reference Fermi-density calculation gives, in relativistic units,
\begin{align}
\Delta E_{\rm HFS}(^{13}\mathrm C)
&=
3.0071\times10^{-6},
\\
\Delta E_{\rm HFS}(^{15}\mathrm N)
&=
-1.8701\times10^{-6}.
\end{align}
Their absolute values correspond to
\begin{align}
\lambda_{\rm HFS}(^{13}\mathrm C)&=807~{\rm nm},
\\
\lambda_{\rm HFS}(^{15}\mathrm N)&=1297~{\rm nm}.
\end{align}
These wavelengths should therefore be regarded as reference values within the present point-magnetic-dipole model rather than as precision spectroscopic predictions; variations of the nuclear-density profile and finite magnetization can alter them. For comparison, the simple point-nucleus Fermi estimate of Eq.~(\ref{eq:HFS_Fermi}) gives the wavelengths listed in Table~\ref{tab:HFS_lambda}, which are approximately $9\%$ and $12\%$ shorter than the reference finite-nucleus results for $^{13}\mathrm C$ and $^{15}\mathrm N$, respectively.

The muonic, nuclear, and total reduced magnetic-dipole matrix elements between two hyperfine sublevels of the $2s$ state are
\begin{align}
^{13}\mathrm C:\quad
\langle F=1\Vert\boldsymbol{\mu}_\mu\Vert
F=0\rangle
&=
-4.1877\times10^{-3} |e|,
\nonumber\\
\langle F=1\Vert\boldsymbol{\mu}_{\rm nuc}\Vert
F=0\rangle
&=
-3.3129\times10^{-4} |e|,
\nonumber\\
\langle F=1\Vert\boldsymbol{\mu}\Vert
F=0\rangle
&=
-4.5190\times10^{-3}|e|,
\\
^{15}\mathrm N:\quad
\langle F=1\Vert\boldsymbol{\mu}_\mu\Vert
F=0\rangle
&=
-4.1875\times10^{-3}|e|,
\nonumber\\
\langle F=1\Vert\boldsymbol{\mu}_{\rm nuc}\Vert
F=0\rangle
&=
+1.3357\times10^{-4}|e|,
\nonumber\\
\langle F=1\Vert\boldsymbol{\mu}\Vert
F=0\rangle
&=
-4.0539\times10^{-3}|e|.
\end{align}

\subsection{PNC electric-dipole amplitudes}

For the ordered $F=0\to F=1$ transition, the amplitudes obtained from Eqs.~(\ref{eq:SI_PNC3}) and (\ref{eq:SD_PNC2}) are written as
\begin{align}
\langle d_{\rm PNC}^{\rm SI}\rangle
&=
G_{\rm F} Q_W |e|\,D_{\rm SI},
\\
\langle d_{\rm PNC}^{\rm SD}\rangle
&=
G_{\rm F}\kappa |e|\,D_{\rm SD}.
\end{align}
Here $D_{\rm SI}$ and $D_{\rm SD}$ are the corresponding dimensionless atomic coefficients; all dependence on the weak couplings is factored into $Q_W$ and $\kappa$, respectively. The individual contributions of the lowest intermediate bound states are listed in Table~\ref{tab:pnc_contributions}. The ``$n=4$--$10$'' rows contain the corresponding sums over this range of principal quantum numbers.

\begin{table*}[t]
\caption{
Contributions of the intermediate $np_j$ states to the PNC-induced electric-dipole amplitudes. The coefficients are defined by
$\langle d_{\rm PNC}^{\rm SI}\rangle
 =G_{\rm F}Q_W|e|D_{\rm SI}$ and
$\langle d_{\rm PNC}^{\rm SD}\rangle
 =G_{\rm F}\kappa |e| D_{\rm SD}$.
The totals are evaluated from the unrounded contributions.
}
\label{tab:pnc_contributions}
\begin{ruledtabular}
\begin{tabular}{c c r r}
Isotope &
Intermediate state &
$D_{\rm SI}$ &
$D_{\rm SD}$
\\
\hline
\noalign{\vskip 0.4ex}
$^{13}\mathrm C$
& $2p_{1/2}$
& $ 4.36053\times10^{1}$
& $-9.54326\times10^{3}$ \\
& $2p_{3/2}$
& $0$
& $ 1.35394\times10^{2}$ \\
& $3p_{1/2}$
& $-7.87211\times10^{-5}$
& $-6.31762$ \\
& $3p_{3/2}$
& $0$
& $ 5.25233\times10^{-2}$ \\
& $n=4$--$10,\ p_{1/2}$
& $-2.26383\times10^{-5}$
& $-2.42367$ \\
& $n=4$--$10,\ p_{3/2}$
& $0$
& $ 2.01719\times10^{-2}$ \\
& Total
& $4.36052\times 10^1$
& $-9.41653\times10^{3}$ \\
\hline
\noalign{\vskip 0.4ex}
$^{15}\mathrm N$
& $2p_{1/2}$
& $-5.83901$
& $-5.48507\times10^{3}$ \\
& $2p_{3/2}$
& $0$
& $ 8.39546\times10^{1}$ \\
& $3p_{1/2}$
& $ 4.10221\times10^{-5}$
& $-7.20272$ \\
& $3p_{3/2}$
& $0$
& $ 7.46814\times10^{-2}$ \\
& $n=4$--$10,\ p_{1/2}$
& $ 1.17991\times10^{-5}$
& $-2.75728$ \\
& $n=4$--$10,\ p_{3/2}$
& $0$
& $ 2.86328\times10^{-2}$ \\
& Total
& $-5.83896$
& $-5.41097\times10^{3}$
\end{tabular}
\end{ruledtabular}
\end{table*}

The $2p_{1/2}$ state clearly dominates both PNC amplitudes. For the SI amplitude the relative contribution of all explicitly included states with $n\geq3$ is below $10^{-5}$
in both systems. For the SD amplitude, the combined contribution of the $n=3$ states and of the $n=4$--$10$ tail is below $2\times10^{-3}$ of the total result in both
systems.

The independent finite-basis calculation, including both continuum sectors, confirms the convergence of the intermediate-state sum. The contribution beyond the explicitly evaluated $n\leq10$ bound states changes the SD amplitude by only about $0.04\%$ for $^{13}\mathrm C$ and $0.08\%$ for $^{15}\mathrm N$, while its effect on the SI amplitude is below $2\times10^{-4}\%$ in both systems.

\am{The effect of intermediate-state radiative widths is small for the dominant $2p_{1/2}$ contribution. Leading hydrogenic estimates give energy widths $\gamma_{2p} \simeq 0.12~\mathrm{eV}$ and $0.22~\mathrm{eV}$ for carbon and nitrogen, respectively, compared with $2s$--$2p_{1/2}$ separations of $26$--$27~\mathrm{eV}$ and
$70$--$71~\mathrm{eV}$. Replacing $\Delta=E_{2s}-E_{2p}$ by $\Delta+i\gamma_{2p}/2$ gives
\begin{equation}
\operatorname{Re} \, \frac{1}{\Delta+i\gamma_{2p}/2}
=\frac{\Delta}{\Delta^2+\gamma_{2p}^2/4}.
\label{eq:real_denominator_width}
\end{equation}
The relative correction to the dispersive part is therefore quadratic in
$\gamma_{2p}/(2|\Delta|)<3\times10^{-3}$. It remains below $2\times10^{-5}$ also for the difference of denominators entering the SI contribution. These corrections are negligible at the accuracy of the present calculation of the dispersive
PNC coefficients.}

\subsection{Degree of parity nonconservation}

We represent the PNC degree as
\begin{equation}
\mathcal P
=
C_{Q_W}Q_W
+
C_\kappa\kappa.
\label{eq:P_calculated}
\end{equation}
Here $C_{Q_W}$ and $C_\kappa$ are the atomic coefficients multiplying the SI and SD weak couplings $Q_W$ and $\kappa$, respectively. Using the Fermi distribution as the reference, we obtain
\begin{align}
\mathcal P(^{13}\mathrm C)
&=
-5.9\times10^{-8}Q_W
+
1.3\times10^{-5}\kappa,
\label{eq:pnc_C13}
\\
\mathcal P(^{15}\mathrm N)
&=
8.8\times10^{-9}Q_W
+
8.1\times10^{-6}\kappa.
\label{eq:pnc_N15}
\end{align}
The coefficients above refer to the ordered $F=0\to F=1$ transition. Since the lower hyperfine level in $^{15}{\rm N}$ has $F=1$, the PNC degree relevant to
absorption is $\mathcal P_{\rm abs}(^{15}{\rm N})=-\mathcal P(^{15}{\rm N})$.
For $^{13}{\rm C}$ one has $\mathcal P_{\rm abs}=\mathcal P$. The atomic coefficients obtained with the two nuclear-density models are summarized in Table~\ref{tab:density_model_dependence}.

\begin{table}[t]
\caption{
Dependence of the final atomic coefficients on the nuclear density model for the ordered $F=0\to F=1$ transition. \am{The $\delta_{\rm shape}$ rows give the relative changes with respect to the Fermi distribution, expressed in percent.}
}
\label{tab:density_model_dependence}
\begin{ruledtabular}
\begin{tabular}{c c r r}
Isotope & Model & $10^8C_{Q_W}$ & $10^5C_\kappa$\\
\hline
\noalign{\vskip 0.4ex}
$^{13}\mathrm C$ & sphere & $-5.66809$ & $1.24964$\\
                 & Fermi  & $-5.87767$ & $1.26928$\\
                 & \am{$\delta_{\rm shape}$}
                 & \am{$3.57\%$} & \am{$1.55\%$}\\[0.4ex]
$^{15}\mathrm N$ & sphere & $0.848452$ & $0.802050$\\
                 & Fermi  & $0.877348$ & $0.813039$\\
                 & \am{$\delta_{\rm shape}$}
                 & \am{$3.29\%$} & \am{$1.35\%$}
\end{tabular}
\end{ruledtabular}
\end{table}

\am{Taking the Fermi distribution as the reference, we define the relative change of either atomic coefficient $C$ as $\delta_{\rm shape}(C)=|C^{\rm sph}-C^{\rm F}|/|C^{\rm F}|$. The corresponding relative changes are listed in Table~\ref{tab:density_model_dependence}.} These differences characterize the sensitivity to the common radial profile used for the charge, weak-charge, and spin-dependent densities. They do not account for possible differences among the corresponding physical distributions. The Fermi--sphere comparison is performed at fixed input $r_{\rm rms}$. The sensitivity to the experimental uncertainty of the charge radius was estimated independently by increasing $r_{\rm rms}$ by one quoted standard uncertainty, $0.0045~{\rm fm}$ for $^{13}\mathrm C$ and $0.0080~{\rm fm}$ for $^{15}\mathrm N$. In the auxiliary finite-basis calculation this changes $C_{Q_W}$ by $1.4\%$ and $1.8\%$, respectively, and $C_\kappa$ by $0.7\%$ and $0.9\%$. Thus, the radius uncertainty produces a somewhat smaller, but comparable-in-scale, sensitivity than the variation of the nuclear-density profile.

\am{For the present transition, the diagonal weak--hyperfine contribution can be characterized by the equivalent coupling
\begin{equation}
\kappa_{\rm hf}^{\rm diag}=Q_W\frac{C_{Q_W}}{C_\kappa}.
\label{eq:kappa_hf_diag}
\end{equation}
Using the Fermi-model coefficients and the adopted weak charges gives
$\kappa_{\rm hf}^{\rm diag}=+0.0302$ for $^{13}\mathrm C$ and
$-0.00803$ for $^{15}\mathrm N$. For the values of $\kappa$ listed in
Table~\ref{tab:kappa_models}, the diagonal weak--hyperfine contribution has the opposite sign to the direct SD term in both ions and is comparable
to it in $^{13}\mathrm C$, leading to destructive interference in the calculated PNC signal. The comparison of the anapole and neutral-current contributions discussed above refers specifically to the direct SD sector. The diagonal weak--hyperfine contribution is already contained in $C_{Q_W}Q_W$ and is therefore
kept separate from $\kappa=\kappa_a+\kappa_Z$.}

The main omitted contributions are the finite-magnetization correction, hyperfine-induced wave-function corrections (off-diagonal hyperfine mixing), higher-order radiative and recoil effects, and nuclear polarization. Nuclear-polarization corrections to the hyperfine structure of muonic atoms have recently been investigated explicitly in heavy systems~\cite{Vandeleur2025}; no corresponding isotope-specific correction is included here. The Bohr--Weisskopf correction may be particularly relevant to $C_{Q_W}$ because the contribution generated by the SI PNC Hamiltonian depends on differences of hyperfine-resolved energy denominators. Since the purpose of the present work is to establish the scale and system dependence of the effect rather than to provide a precision prediction, we quote the conservative working allowances
\begin{equation}
\frac{\delta C_{Q_W}}
{\left|C_{Q_W}\right|}
=
10\%,
\qquad
\frac{\delta C_\kappa}{|C_\kappa|}
=
5\%.
\label{eq:working_allowances}
\end{equation}
These values are phenomenological estimates rather than statistically defined uncertainties. In particular, the allowance assigned to $C_{Q_W}$ does not account for the omitted hyperfine-induced corrections to the wave functions.

For the nuclear-model values of $\kappa$ listed in Table~\ref{tab:kappa_models}, the SI and SD contributions to the absorption asymmetry have opposite signs. Their magnitudes are comparable in $^{13}{\rm C}$, whereas the SD contribution is larger in $^{15}{\rm N}$. 

Within the SD sector, direct $Z^0$-boson exchange is the largest contribution to $\kappa$ in $^{13}{\rm C}$ in the single-particle model and remains comparable to the anapole contribution in the NCS model calculation; in $^{15}{\rm N}$, the interaction is anapole dominated in both models. We therefore quote the two atomic coefficients separately rather than interpret their sum as a final Standard-Model prediction. The larger $C_{Q_W}$ in $^{13}{\rm C}$ originates mainly from its smaller $2s_{1/2}$--$2p_{1/2}$ denominator, while the larger predicted $\kappa$ in $^{15}{\rm N}$ can make
the magnitude of its calculated PNC effect comparable to or larger than that in $^{13}{\rm C}$.

\section{Cavity-enhanced experimental prospects} \label{sec:experiment}

In this section we present a simplified theoretical model of a possible cavity-enhanced polarimetric experiment and estimate the conditions under which the calculated PNC amplitude could be observed by measuring optical rotation on the $2s_{1/2}$ hyperfine transition. The aim is to identify the characteristic optical and column-density requirements and to determine the scale of the principal experimental limitations. We do not attempt to specify a complete experimental apparatus. In this section, $\hbar$ and $c$ are restored explicitly whenever laboratory units are used.

\subsection{Geometry and characteristic time scales}

We consider a beam of metastable hydrogen-like muonic ions crossing a linearly polarized laser mode inside an optical cavity. Let $d$ be the dimension of the ion beam along the direction of laser propagation and $2w_L$ the laser-mode diameter along the direction of motion of the ions. The effective optical path and the transit time are
\begin{equation}
l_{\rm eff}=N_{\rm eff}d,
\qquad
\tau_{\rm tr}=\frac{2w_L}{v_b},
\label{eq:experimental_geometry}
\end{equation}
where $N_{\rm eff}$ is the effective number of passages through the ion beam and $v_b$ is the beam velocity. We use
\begin{equation}
d=3~{\rm cm},
\qquad
2w_L=1~{\rm mm}
\label{eq:beam_dimensions}
\end{equation}
as representative geometrical parameters.  For the systems considered here we adopt
\begin{equation}
\tau_{2s}=10^{-8}~{\rm s}.
\label{eq:tau_2s_benchmark}
\end{equation}
This lifetime determines the homogeneous contribution to the line width,
\begin{equation}
\Gamma_{\rm hom}\sim\tau_{2s}^{-1}
=10^8~{\rm s}^{-1}.
\label{eq:homogeneous_width}
\end{equation}
A transit time not exceeding the lifetime requires
\begin{equation}
v_b\gtrsim
\frac{2w_L}{\tau_{2s}}
\simeq10^5~{\rm m/s}.
\label{eq:required_velocity}
\end{equation}
This velocity corresponds to kinetic energies of approximately $0.7~{\rm keV}$ for $^{13}\mathrm C$ and $0.8~{\rm keV}$ for $^{15}\mathrm N$. It is therefore kinematically accessible for an ion beam, although the short lifetime requires the metastable ions to be produced and accelerated very close to the interaction region.

For a passive Fabry--Perot cavity, finesse values above $10^6$ have been demonstrated using mirrors with sub-ppm excess losses \cite{Jin2022}. Accordingly, $N_{\rm eff}=10^5$--$10^6$ represents an experimentally motivated range for a passive high-finesse cavity. Much larger effective absorption paths have been achieved using active intracavity absorption spectroscopy. In particular, effective paths up to $7\times10^4~{\rm km}$ were reported for a continuous-wave dye laser \cite{Baev1999}. In this case the enhancement results from the spectral dynamics of a multimode laser with gain and is not identical to the geometrical number of round trips in a passive cavity. We therefore include $N_{\rm eff}=10^7$ only as an exploratory ICAS-equivalent scenario. Its use for polarization rotation would require a dedicated treatment of the polarization eigenmodes, gain dynamics, and intracavity retarder losses.

In a linear two-mirror cavity, the PNC rotation acquired on the backward pass would cancel the forward-pass contribution unless the polarization state is transformed at reflection. A suitable arrangement of intracavity quarter-wave
plates transforms the polarization at each reflection so that the PNC contributions add over a complete round trip. This can be verified directly using the Jones-matrix formalism.

\subsection{Line profile, rotation, and absorption}

We introduce the real-valued absorptive and dispersive profile functions $f(u,q)$ and $g(u,q)$ through~\cite{Khriplovich1991}
\begin{equation}
f(u,q)+ig(u,q)
=
\sqrt{\pi}\,w(u+iq),
\label{eq:complex_Voigt_exp}
\end{equation}
where $w(z)$ is the Faddeeva function,
\begin{equation}
w(z)
=
\mathrm{e}^{-z^2}\operatorname{erfc}(-iz).
\label{eq:Faddeeva_function}
\end{equation}
The dimensionless detuning and homogeneous-width parameters are
\begin{equation}
u=\frac{\omega_L-\omega_0}{D},
\qquad
q=\frac{\Gamma_{\rm hom}}{2D},
\label{eq:detuning_exp}
\end{equation}
where
\begin{equation}
D
=
\omega_0
\left(
\frac{2k_{\rm B}T_\perp}{Mc^2}
\right)^{1/2}
\label{eq:Doppler_exp}
\end{equation}
is the transverse Doppler width. The absorption coefficient is written as
\begin{equation}
 \alpha(u,q)=n_g\,\sigma_0 f(u,q),
\label{eq:absorption_coefficient}
\end{equation}
where $n_g$ is the number density of ions in the initial hyperfine state. The characteristic cross section can be expressed directly through the total reduced magnetic-dipole matrix element of Eq.~(\ref{eq:M1_total_reduced}) as~\cite{Sobelman1992,Hilborn1982}
\begin{equation}
 \sigma_0
 =
 \frac{\omega_0}
 {3(2F_g+1)D}
 \left|
 \left\langle
 2s_{1/2}F_e
 \left\Vert
 \boldsymbol{\mu}
 \right\Vert
 2s_{1/2}F_g
 \right\rangle
 \right|^2 .
\label{eq:sigma0_experiment}
\end{equation}

Let
\begin{equation}
 x\equiv n_g l_{\rm eff}
 =n_g dN_{\rm eff}
\label{eq:column_density_definition}
\end{equation}
be the effective column density of ions in the initial hyperfine state. The dispersive PNC rotation angle and the transmission are then given by
\begin{align}
 \psi(u)
 &=
 \frac{\mathcal P}{2}\,
 x\sigma_0 g(u,q),
\label{eq:PNC_rotation_experiment}
\\
 \mathcal T(u)
 &=
 \exp\left[-x\sigma_0 f(u,q)\right].
\label{eq:transmission_experiment}
\end{align}
Here $\mathcal P$ refers to the actual absorption transition $F_g\to F_e$; therefore, for $^{15}\mathrm N$ its sign is opposite to that of the $0\to1$ coefficient quoted in Eq.~(\ref{eq:pnc_N15}). Only $|\mathcal P|$ enters the feasibility estimates below. For the order-of-magnitude estimates we assume an effective transverse temperature
\begin{equation}
 T_\perp=1~{\rm K}.
\end{equation}
The magnetic-dipole transition rates in the two systems are both of order $10^{-3}~{\rm s}^{-1}$. However, since the absorption cross section also depends on the transition frequency, the Doppler width, and the statistical weight of the initial hyperfine level, we do not average the cross sections of the two transitions.

The $^{13}{\rm C}$ transition provides the more favorable benchmark. For this system,
\begin{equation}
 D\simeq2.8\times10^8~{\rm s}^{-1},
 \qquad
 \sigma_0\simeq1.0\times10^{-20}~{\rm cm^2}.
\label{eq:C_optical_parameters}
\end{equation}
The dispersive profile reaches its maximum at a detuning of approximately one transverse Doppler width, $|u|\simeq1$, where both the absorptive and dispersive profile functions are of order unity.

\subsection{Signal and photon shot noise} \label{subsec:shot_noise}

Following the approach used in Refs.~\cite{Chubukov2019,Chubukov2021,Chubukov2023},
we describe the polarimetric measurement in terms of the rotation signal and
the fundamental photon shot noise. For $N_\gamma$ registered photons, the shot-noise-limited signal-to-noise ratio is
\begin{equation}
\frac{R}{F}
=
|\psi(u)|\sqrt{N_\gamma},
\qquad
\frac{R}{F}>1
\end{equation}
for an observable rotation signal.

The intensity reaching the detector is
\begin{equation}
I_{\rm out}(u)
=
\delta\,I_{\rm cav}\mathcal T(u),
\end{equation}
where $\delta$ is the effective output-coupling fraction. For an output coupling of the same order as the total round-trip loss, we use the estimate $\delta\sim N_{\rm eff}^{-1}$. Thus the range $N_{\rm eff}=10^5$--$10^7$ corresponds to $\delta=10^{-5}$--$10^{-7}$, up to numerical factors of order unity. The number of photons registered during a measurement time $T_{\rm run}$ is then
\begin{equation}
 N_\gamma(u)
 =
 \frac{
 \delta\,I_{\rm cav}A_LT_{\rm run}
 }{\hbar\omega_0}
 \mathcal T(u),
\label{eq:number_of_photons}
\end{equation}
where $A_L$ is the cross-sectional area of the laser mode. The intracavity intensity determines the interaction with the ions, whereas the product $\delta\,I_{\rm cav}$ determines the photon flux at the detector. 

Using Eqs.~(\ref{eq:PNC_rotation_experiment}) and (\ref{eq:transmission_experiment}), the signal-to-noise ratio takes the form
\begin{align}
\frac{R}{F}
&=
\frac{|\mathcal P|}{2}
x\sigma_0|g(u,q)|
\left(
\frac{\delta\,I_{\rm cav}A_LT_{\rm run}}
{\hbar\omega_0}
\right)^{1/2}
\nonumber \\
{}&\times
\exp\left[
-\frac{x\sigma_0f(u,q)}{2}
\right].
\label{eq:signal_to_noise_full}
\end{align}
In the parameter range considered below, the optical depth is much smaller than unity. Absorption therefore has a negligible effect on the photon statistics, and
Eq.~(\ref{eq:signal_to_noise_full}) reduces to
\begin{equation}
 \frac{R}{F}
 \simeq
 \frac{|\mathcal P|}{2}\,
 x\sigma_0|g(u,q)|
 \left(
 \frac{
 \delta\,I_{\rm cav}A_LT_{\rm run}
 }{\hbar\omega_0}
 \right)^{1/2}.
\label{eq:signal_to_noise_thin}
\end{equation}

\subsection{Transit-time bleaching and saturation}
\label{subsec:bleaching}

The intracavity intensity is limited by the requirement that the population
of the initial hyperfine component should not be appreciably depleted during
its passage through the laser mode. In the weak-excitation regime, the
absorption probability can be estimated as the product of the photon flux,
the Doppler-averaged absorption cross section, and the transit time
\cite{Hilborn1982}:
\begin{equation}
W_{\rm abs}(u)
\simeq
\frac{I_{\rm cav}\tau_{\rm tr}}{\hbar\omega_0}
\sigma_0 f(u,q).
\label{eq:absorption_probability}
\end{equation}
 In the numerical estimates below we use $\tau_{\rm tr}=10^{-8}~{\rm s}$. Since this
value is comparable to the lifetime $\tau_{2s}$, neglecting the decay of
metastable ions during the transit slightly overestimates the depletion and
therefore gives a conservative estimate. A more complete finite-interaction-time treatment may modify the numerical result by a factor of order unity
\cite{Chubukov2023}.

We define the maximum usable intensity by requiring that the absorption
probability remain below $10\%$,
\begin{equation}
W_{\rm abs}(u) \leq p,
\qquad p=0.1.
\label{eq:bleaching_condition}
\end{equation}
This gives
\begin{equation}
I_{\max}(u)
=
\frac{p\hbar\omega_0}
{\tau_{\rm tr}\sigma_0 f(u,q)}.
\label{eq:maximum_intensity}
\end{equation}
Then, one obtains
\begin{equation}
I_{\max}\sim4\times10^8~{\rm W/cm^2}.
\label{eq:maximum_intensity_numeric}
\end{equation}

It is useful to distinguish this bleaching limit from the usual saturation intensity. A characteristic saturation intensity can be estimated by requiring the stimulated absorption rate to become comparable to the relaxation rate of the $2s$ state,
\begin{equation}
\frac{I_{\rm sat}\sigma_0 f(u,q)}{\hbar\omega_0}
\sim
\frac{1}{\tau_{2s}}.
\end{equation}
Thus,
\begin{equation}
I_{\rm sat}(u)
\sim
\frac{\hbar\omega_0}{\tau_{2s}\sigma_0 f(u,q)}
\sim 4\times10^9~{\rm W/cm^2}.
\label{eq:saturation_intensity}
\end{equation}
Since $\tau_{\rm tr}\sim\tau_{2s}$, the intensity defined by the $10\%$ depletion condition is approximately $I_{\max}\sim 0.1 I_{\rm sat}$. The two estimates differ mainly by the chosen small-depletion parameter $p$.

As an optimistic theoretical benchmark, we use
\begin{equation}
I_{\rm cav}=10^8~{\rm W/cm^2}.
\label{eq:intracavity_intensity_benchmark}
\end{equation}
The transition therefore remains well below saturation, and the linear-response treatment used for the rotation and absorption is self-consistent.

\am{
The cavity field also couples $2s$ off-resonantly to the $2p_j$ states, producing an ac Stark shift and additional quenching. These effects are suppressed by the large $2s$--$2p_j$ energy separations, of the order of tens of electronvolts. At $I_{\rm cav}=10^8~\mathrm{W/cm^2}$, a conservative second-order estimate including both $2p_{1/2}$ and
$2p_{3/2}$ and the maximum standing-wave field gives
\begin{equation}
|\delta\omega_{\rm HFS}|\lesssim10^4~\mathrm{s^{-1}},
\qquad
\Gamma_{\rm ind}\lesssim10^2~\mathrm{s^{-1}}.
\end{equation}
The hyperfine-transition shift is much smaller than $\Gamma_{\rm hom}\sim10^8~\mathrm{s^{-1}}$, and the additional quenching depletes the $2s$ population by less than one part in a million during
$\tau_{\rm tr}=10^{-8}~\mathrm{s}$. For the unpolarized ensemble of ions considered here, these effects are helicity even. A detailed analysis of experimental systematics depends on the specific cavity and detection protocol and lies beyond the scope of this proposal; signal-reversal and background-rejection methods for cavity-enhanced PNC polarimetry are discussed in Refs.~\cite{Bougas2012,Bougas2014}.
}

\subsection{Required column density} \label{subsec:required_density}

For the numerical estimate we take
\begin{align}
 |\mathcal P| &= 10^{-7},
 &
 I_{\rm cav} &= 10^8~{\rm W/cm^2},
 \nonumber\\
 A_L &= 10^{-2}~{\rm cm^2},
 &
 T_{\rm run} &= 10^6~{\rm s}.
\end{align}
The value $|\mathcal P|=10^{-7}$ is used as a representative normalization rather than as a unique nuclear prediction. The value $T_{\rm run}=10^6~{\rm s}$ represents an
accumulated live time during which metastable ions are present in the interaction region. For the central cavity parameters
\begin{equation}
 N_{\rm eff}=10^6,
 \qquad
 \delta=10^{-6},
\label{eq:central_cavity_parameters}
\end{equation}
the number of registered photons is
\begin{equation}
 N_\gamma
 \simeq
 \frac{
 \delta\,I_{\rm cav}A_LT_{\rm run}
 }{\hbar\omega_0}
 \simeq4\times10^{24},
\label{eq:number_of_photons_numeric}
\end{equation}
where the very small resonant absorption has been neglected.

The condition $R/F=1$ gives the required effective column density
\begin{equation}
 x_{\rm req}
 =
 \frac{2}
 {|\mathcal P|\sigma_0|g(u,q)|}
 \left(
 \frac{\hbar\omega_0}
 {\delta\,I_{\rm cav}A_LT_{\rm run}}
 \right)^{1/2}.
\label{eq:required_column_density}
\end{equation}
Using
\begin{align}
 \sigma_0 &\simeq 1.0\times10^{-20}~{\rm cm^2}, \\
 |g(u,q)| &\simeq1,
 \qquad
 |u|\simeq1,
\end{align}
we obtain for the central benchmark
\begin{equation}
 x_{\rm req}
 =
 n_g dN_{\rm eff}
 \sim 1.5\times10^{15}~{\rm cm^{-2}}.
\label{eq:required_column_density_numeric}
\end{equation}
For a particle-beam width $d=3~{\rm cm}$, the corresponding local density is
\begin{equation}
 n_g^{\rm req}
 =
 \frac{x_{\rm req}}{dN_{\rm eff}}.
\label{eq:required_local_density_general}
\end{equation}
Representative values are listed in
Table~\ref{tab:required_density_delta}.

\begin{table}[t]
\caption{
Required effective column density and local density for
$I_{\rm cav}=10^8~{\rm W/cm^2}$,
$A_L=10^{-2}~{\rm cm^2}$,
$T_{\rm run}=10^6~{\rm s}$, and
$|\mathcal P|=10^{-7}$. The estimate
$\delta=N_{\rm eff}^{-1}$ is used.
}
\label{tab:required_density_delta}
\begin{ruledtabular}
\begin{tabular}{c c c c}
$N_{\rm eff}$ &
$\delta$ &
$x_{\rm req}$ (${\rm cm^{-2}}$) &
$n_g^{\rm req}$ (${\rm cm^{-3}}$)
\\
\hline
\noalign{\vskip 0.4ex}
$10^5$ & $10^{-5}$ & $4\times10^{14}$ & $1\times10^9$ \\
$10^6$ & $10^{-6}$ & $1.5\times10^{15}$ & $5\times10^8$ \\
$10^7$ & $10^{-7}$ & $4\times10^{15}$ & $1\times10^8$
\end{tabular}
\end{ruledtabular}
\end{table}

When $\delta\simeq N_{\rm eff}^{-1}$, increasing the number of passages enhances the rotation linearly but reduces the detected photon flux. As a result,
\begin{equation}
 n_g^{\rm req}\propto N_{\rm eff}^{-1/2},
\label{eq:density_Neff_scaling}
\end{equation}
rather than $N_{\rm eff}^{-1}$.

At the required column density, the optical depth is only $x_{\rm req}\sigma_0f(u,q)\sim10^{-5}$, so that $\mathcal T\simeq0.99999$. An optical depth of order unity would require $x\sim[\sigma_0f(u,q)]^{-1}\sim1.5\times10^{20}~{\rm cm^{-2}}$, approximately five orders of magnitude above
$x_{\rm req}$.

Continuous neutral-atom beams with densities of order $10^{10}~{\rm cm^{-3}}$ have been demonstrated~\cite{DeGraffenreid2000}, and pulsed supersonic beams can reach much larger
peak densities \cite{Catanese2018}. The densities in Table~\ref{tab:required_density_delta} are therefore not exceptional from the
general perspective of atomic-beam spectroscopy. The essential difficulty is
to obtain the same density in the required metastable state of a hydrogen-like muonic ion.

\subsection{Source-rate limitation} \label{subsec:muonic_sources}

Negative-muon beams with rates of order $10^7$--$10^8~{\rm s^{-1}}$ are available at high-intensity facilities~\cite{Cook2017}. However, muonic atoms are normally formed by stopping the muons in a target, and no source delivering a collimated beam of metastable hydrogen-like muonic carbon or nitrogen ions has yet been demonstrated. Attempts to extract a muonic-atom beam have so far yielded no conclusive signal~\cite{Yoshida2018}.

The source requirement can be illustrated using the central benchmark of Table~\ref{tab:required_density_delta}. The number of metastable ions that must be present simultaneously in the interaction region is
\begin{equation}
N_{\rm req}
\sim
n_g^{\rm req} V_{\rm int}
\sim
10^7,
\end{equation}
where $V_{\rm int}\sim3\times10^{-2}~{\rm cm^3}$. Since the ions cross the laser mode in approximately $\tau_{\rm tr}=10^{-8}~{\rm s}$, a continuous implementation
would require a replenishment rate of order
\begin{equation}
\dot N_{2s}^{\rm req}
\sim
\frac{N_{\rm req}}{\tau_{\rm tr}}
\sim
10^{15}~{\rm s^{-1}}.
\end{equation}

This rate exceeds present primary negative-muon rates by more than seven orders of magnitude even before accounting for the fraction of captures populating the $2s$ state, the required ionic charge state, extraction and transport losses, electron
refilling, and geometrical acceptance. Pulsed sources may provide a substantially larger instantaneous population, but their assessment requires a dedicated treatment of the pulse structure, duty factor, ion extraction, and synchronization with the optical
cavity. The production of a sufficiently intense beam of metastable hydrogen-like muonic ions therefore remains the main experimental challenge. Within the simplified optical model, photon shot noise, resonant absorption, and atomic bleaching do not impose a fundamental limitation.

\section{Conclusion} \label{sec:conclusion}

We have studied parity nonconservation in the $2s_{1/2}$ hyperfine transition of hydrogen-like muonic $^{13}\mathrm C$ and $^{15}\mathrm N$. Within the
diagonal-hyperfine approximation, the calculated contributions to the absorption asymmetry are
\begin{align}
\mathcal P_{\rm abs}(^{13}\mathrm C)
&=
-5.9\times10^{-8}Q_W
+
1.3\times10^{-5}\kappa,
\\
\mathcal P_{\rm abs}(^{15}\mathrm N)
&=
-8.8\times10^{-9}Q_W
-
8.1\times10^{-6}\kappa.
\end{align}
The $Q_W$ terms are generated by the nuclear-spin-independent (SI) PNC Hamiltonian and are evaluated with hyperfine-resolved energy denominators in the diagonal-hyperfine approximation. The terms proportional to $\kappa=\kappa_a+\kappa_Z$ are generated by the nuclear-spin-dependent PNC Hamiltonian. Hyperfine-induced corrections to the wave functions are not included in the present calculation.

Both calculated contributions are dominated by the $2p_{1/2}$ intermediate state. The Fermi--sphere comparison changes $C_{Q_W}$ by $3.3$--$3.6\%$ and $C_\kappa$ by $1.4$--$1.5\%$. Within the spin-dependent sector, $^{13}\mathrm C$ has enhanced relative sensitivity to direct $Z^0$-boson exchange, while $^{15}\mathrm N$ is expected to be predominantly anapole sensitive.

An idealized cavity-polarimetry model gives a required column density of order
$10^{15}~{\rm cm^{-2}}$. The main experimental limitation is the production and replenishment of a sufficiently dense beam of metastable hydrogen-like muonic ions. The present work therefore establishes the scale of the effect and identifies the principal requirements for a possible future measurement, rather than providing a precision prediction or a complete experimental design.


\begin{acknowledgments}
This work was supported by the Foundation for the Advancement of Theoretical Physics and Mathematics BASIS (Project No.~25-1-3-40-1).
\end{acknowledgments}


\section*{Data Availability}

The numerical data that support the findings of this paper are available from the corresponding author upon reasonable request.

\begin{widetext}

\am{ \appendix

\section{Angular reduction of the PNC and M1 amplitudes}
\label{app:angular}

Detailed expressions for the reduced matrix elements of the SI and SD parity-violating E1 amplitudes can be found in Refs.~\cite{Porsev01,Johnson03}. The expressions below are used both in the explicit summation over bound states and in the finite-basis
calculation described in Sec.~\ref{sec:calculations}. For the SI amplitude, we have
\begin{align}
\langle \gamma_fJ_f; F_f || \boldsymbol{d}^{\rm PNC}_{\rm SI} || \gamma_iJ_i; F_i \rangle &= -\frac{G_{\rm F}}{2\sqrt{2}}\, Q_W\,
(-1)^{I+F_i+J_f+1}
\sqrt{(2F_f+1)(2F_i+1)}
\begin{Bmatrix}
J_i & J_f & 1 \\
F_f & F_i & I
\end{Bmatrix}
 \nonumber \\
&{}\times \sum_n\left[
\frac{
\langle \gamma_f J_f ||  \boldsymbol{d} || \gamma_nJ_n\rangle
\langle \gamma_nJ_n | (\gamma_5/i)\, \rho(r) | \gamma_i J_i\rangle
}{
E_i-E_{nJ_nF_i}
}
\right.
\left.
+
\frac{
\langle \gamma_fJ_f | (\gamma_5/i)\, \rho(r) | \gamma_nJ_n\rangle
\langle \gamma_nJ_n ||  \boldsymbol{d} || \gamma_i J_i\rangle
}{
E_f-E_{nJ_nF_f}
}
\right].
\label{eq:SI_PNC}
\end{align}
In our case $J_i=J_f=1/2\equiv J$ and $\gamma_i=\gamma_f\equiv \gamma$. 
The sum runs over odd-parity states with $J_n=1/2$, since the pseudoscalar PNC operator cannot mix different $J$. For a Hermitian irreducible tensor operator $T^{(k)}$ of rank $k$, the reduced matrix elements satisfy the general symmetry relation
\begin{equation}
\langle \gamma_a J_a
||
T^{(k)}
||
\gamma_b J_b
\rangle
=
(-1)^{J_a-J_b}
\left(
\langle
\gamma_b J_b
||
T^{(k)}
||
\gamma_a J_a
\rangle
\right)^{*},
\label{eq:Hermitian_RME}
\end{equation}
Then, since the PNC matrix element is purely imaginary, $\langle a ||  \boldsymbol{d} || b\rangle \langle b | \gamma_5 | a\rangle= -\langle b ||  \boldsymbol{d} || a\rangle \langle a |  \gamma_5 | b\rangle $, and this leads to simplification of Eq.~\eqref{eq:SI_PNC}:
\begin{align}
\langle \gamma J; F_f || \boldsymbol{d}^{\rm PNC}_{\rm SI} || \gamma J; F_i \rangle &= -\frac{G_{\rm F}}{2\sqrt{2}}\, Q_W\,
(-1)^{I+F_i+J+1}
\sqrt{(2F_f+1)(2F_i+1)}
\begin{Bmatrix}
J & J & 1 \\
F_f & F_i & I
\end{Bmatrix}
 \nonumber \\
&{}\times \sum_n
\langle \gamma J ||   \boldsymbol{d} || \gamma_n J_n\rangle
\langle \gamma_nJ_n | (\gamma_5/i)\, \rho(r) | \gamma J\rangle
  \left(\frac{1}{E_i-E_{nJ_nF_i}}  - \frac{1}{E_f-E_{nJ_nF_f}}\right).
\label{eq:SI_PNC2}
\end{align}

For the SD PNC amplitude we have
\begin{align}
\langle \gamma_f J_f; F_f || \boldsymbol{d}_{\rm PNC}^{\rm SD} || \gamma_i J_i;F_i \rangle
&=
-\frac{G_{\rm F}}{\sqrt{2}}
\kappa
\sqrt{
\frac{
(I+1)(2I+1)(2F_i+1)(2F_f+1)
}{
I
}
}
\nonumber\\
&{}\times \sum_n
\Bigg[
(-1)^{J_f-J_i}
\begin{Bmatrix}
J_n & J_i & 1 \\
I & I & F_i
\end{Bmatrix}
\begin{Bmatrix}
J_n & J_f & 1 \\
F_f & F_i & I
\end{Bmatrix}
\frac{
\langle\gamma_f J_f ||  \boldsymbol{d} || \gamma_nJ_n\rangle
\langle \gamma_nJ_n || (\boldsymbol{\alpha}/i) \rho (r)|| \gamma_iJ_i\rangle
}{
E_i-E_{nJ_nF_i}
}
\nonumber\\
&{}+
(-1)^{F_f-F_i}
\begin{Bmatrix}
J_n & J_f & 1 \\
I & I & F_f
\end{Bmatrix}
\begin{Bmatrix}
J_n & J_i & 1 \\
F_i & F_f & I
\end{Bmatrix}
\frac{
\langle \gamma_f J_f || (\boldsymbol{\alpha}/i) \rho (r)|| \gamma_nJ_n\rangle
\langle \gamma_nJ_n || \boldsymbol{d} || \gamma_i J_i\rangle
}{
E_f-E_{nJ_nF_f}
}
\Bigg].
\label{eq:SD_PNC}
\end{align}

For the reduced matrix element of the muonic magnetic dipole operator, the following relation holds:  
\begin{equation}
\langle \gamma_f J_f; F_f \| \boldsymbol{\mu}_{\mu}\| \gamma_i J_i;F_i \rangle = (-1)^{I+F_i+J_f+1}
\sqrt{(2F_f+1)(2F_i+1)}
\begin{Bmatrix}
J_i & J_f & 1 \\
F_f & F_i & I
\end{Bmatrix} 
\langle \gamma_fJ_f \| \boldsymbol{\mu}_{\mu}\| \gamma_iJ_i \rangle 
\end{equation}

\section{Radial matrix elements}
\label{app:radial}

The hyperfine shifts entering the hyperfine-resolved energy denominators in Eqs.~(\ref{eq:SI_PNC3}) and (\ref{eq:SD_PNC2}) are evaluated using the relativistic
Fermi--Breit expression~\cite{BetheSalpeter1957},
\begin{equation}
\Delta E^{\rm FB}_{n\varkappa F} = \frac{\alpha  \varkappa}{ 2IJ(J+1)}\frac{\mu_I}{\mu_N}\frac{1}{m_p}\left[F(F+1)-I(I+1)-J(J+1) \right] \int\limits_0^\infty dr \, r^{-2} G_{n\varkappa}(r) F_{n\varkappa}(r).
\label{eq:FB_shift}
\end{equation}

The bound-muon reduced M1 matrix element entering the denominator of Eq.~(\ref{eq:PNC_degree}) through Eq.~(\ref{eq:M1_total_reduced}) is
\begin{equation}
\langle 2s_{1/2} \|  \boldsymbol{\mu}_\mu\| 2s_{1/2} \rangle =\frac{2\sqrt{6}|e|}{3} \int\limits_0^\infty dr \, r G_{2s} (r) F_{2s}(r)
\label{eq:M1_radial}
\end{equation}
The total matrix element is obtained by adding the nuclear contribution according to Eq.~(\ref{eq:M1_total_reduced}).

The bound-muon reduced E1 matrix element entering Eqs.~(\ref{eq:SI_PNC3}) and (\ref{eq:SD_PNC2}) is
\begin{equation}
\label{eq:E1_radial}
\langle 2s_{1/2} ||  \boldsymbol{d} || \gamma_n J_n\rangle   =  h(J_n) \eta_E |e|  \int\limits_0^\infty dr \, r \left[ F_{2s} (r) F_{n\varkappa_n} (r) + G_{2s}(r) G_{n\varkappa_n} (r) \right]   ,
\end{equation}
where $h(1/2)=2/\sqrt{6}$ and $h(3/2)=2/\sqrt{3}$. Here $\varkappa_n$ denotes the relativistic angular quantum number of the intermediate state: $\varkappa_n=1$ for $np_{1/2}$ and $\varkappa_n=-2$ for $np_{3/2}$. This reduced matrix element enters the
electric-dipole amplitudes between the $2s_{1/2}$ and $np_j$ states.

The SI PNC matrix element entering Eq.~(\ref{eq:SI_PNC3}) is
\begin{equation}
\label{eq:SI_radial}
\langle \gamma_n J_n | (\gamma_5/i) \rho(r) | 2s_{1/2}\rangle  = \int\limits_0^\infty dr \, \rho(r) \left[F_{n\varkappa_n} (r) G_{2s} (r)- G_{n\varkappa_n} (r) F_{2s} (r ) \right].
\end{equation}
The SD PNC matrix element entering Eq.~(\ref{eq:SD_PNC2}) is
\begin{equation}
\label{eq:SD_radial}
\langle \gamma_nJ_n\|(\boldsymbol{\alpha}/i)\rho(r)\|2s_{1/2}\rangle
=
\int\limits_0^\infty dr\,\rho(r)
\left[
A_{J_n}G_{n\varkappa_n}(r)F_{2s}(r)
-
B_{J_n}F_{n\varkappa_n}(r)G_{2s}(r)
\right],
\end{equation}
where $A_{J_n}=\langle \varkappa_n\|\boldsymbol{\sigma}\|1\rangle$ and $B_{J_n}=\langle -\varkappa_n\|\boldsymbol{\sigma}\|-1\rangle$. In turn, the required reduced matrix elements of the Pauli matrix are
\begin{equation}
\langle \varkappa_a\|\boldsymbol{\sigma}\|\varkappa_b\rangle
=
\delta_{l_a l_b}
(-1)^{l_a+1/2+j_a+1}
\sqrt{6(2j_a+1)(2j_b+1)}
\begin{Bmatrix}
1/2 & j_a & l_a\\
j_b & 1/2 & 1
\end{Bmatrix}.
\end{equation}
For the intermediate $np_{1/2}$ state, one obtains
$A_{1/2}
=
\langle 1\|\boldsymbol{\sigma}\|1\rangle
=
-\frac{\sqrt6}{3}$, $B_{1/2}
=
\langle -1\|\boldsymbol{\sigma}\|-1\rangle
=
\sqrt6$. For the intermediate $np_{3/2}$ state,
$A_{3/2}
=
\langle -2\|\boldsymbol{\sigma}\|1\rangle
=
-\frac{4\sqrt3}{3}$, $B_{3/2}
=
\langle 2\|\boldsymbol{\sigma}\|-1\rangle
=
0$.
}
\end{widetext}


\end{document}